\RequirePackage{fix-cm}
\documentclass[smallextended]{svjour3}       
\smartqed  
\usepackage{graphicx}
\usepackage{algorithmic}
\usepackage{algorithm}
\usepackage{subfigure}
\usepackage{natbib}
\usepackage{url}

\newtheorem{assumption}{{\bf Assumption}}

\begin{document}

\title{MYE: Missing Year Estimation in Academic Social Networks}

\author{Tom Z. J. Fu \and Qiufang Ying \and Dah Ming Chiu}

\authorrunning{Fu et al} 

\institute{Tom Z. J. Fu, Qiufang Ying\at
              Room 725, Ho Sin Hang Engineering Building,
              Dept. of Information Engineering,\\
              The Chinese University of Hong Kong, Shatin, N.T. Hong Kong\\
              Tel.: +852-39634296\\
              Fax: +852-26035032\\
              \email{\{fuzhengjia,yingqiufan\}@gmail.com}           
           \and
              Dah Ming Chiu\at
              Room 836, Ho Sin Hang Engineering Building,
              Dept. of Information Engineering,\\
              The Chinese University of Hong Kong, Shatin, N.T. Hong Kong\\
              Tel.: +852-39438357\\
              Fax: +852-26035032\\
              \email{dmchiu@ie.cuhk.edu.hk}\\
              \\
              \emph{
              Dr. Fu has moved to:\\
              Illinois at Singapore Pte Ltd,
              Advanced Digital Sciences Center (ADSC)\\
              1 Fusionopolis Way, \#08-10 Connexis North Tower,
              Singapore 138632\\
              Tel.: +65-65919093\\
              Fax: +65-65919091\\}  
}

\date{Received: date / Accepted: date}

\maketitle

\begin{abstract}
In bibliometrics studies, a common challenge is how to deal with
incorrect or incomplete data. However, given a large volume of data, 
there often exists certain relationships between the data items that
can allow us to recover missing data items and correct erroneous data.
In this paper, we study a particular problem of this sort -
estimating the missing year information associated with publications
(and hence authors' years of active publication). We first propose a
simple algorithm that only makes use of the ``direct'' information,
such as paper citation/reference relationships or paper-author
relationships.
The result of this simple algorithm is used as a benchmark for
comparison. Our goal is to develop algorithms that increase both the
coverage (the percentage of missing year papers recovered) and
accuracy (mean absolute error of the estimated year to the real
year). We propose some advanced algorithms that extend inference by
information propagation.
For each algorithm, we propose three versions according to the given
academic social network type: a) Homogeneous (only contains paper
citation links), b) Bipartite (only contains paper-author
relations), and, c) Heterogeneous (both paper citation and
paper-author relations). We carry out experiments on the three
public data sets (MSR Libra, DBLP and APS), and evaluated them by
applying the K-fold cross validation method. We show that the
advanced algorithms can improve both coverage and accuracy.
\keywords{Data Cleaning \and Academic Social Network \and Paper
Citation} 
\end{abstract}

\section{Introduction}
Academic publication analysis has always been of interest to the
research community. Earlier focus includes citation analysis and
journal impact factor analysis, to help evaluate research impact. In
recent years, there has been an increasing interest in the social aspects of
this research, for example, there exist studies of patterns of
collaborations, automatically inferring advisor-advisee
relationships, and finding or predicting leaders and rising stars in
research areas. A common challenge to such research is how to deal
with the lack of data, or when data is available, its incorrectness
and incompleteness. However, since the data volume is large, and
there exists all kinds of relationships between data items, it is
often possible to recover certain missing (or correct erroneous)
data items from the data we have. In this paper, we study a
particular problem of this sort - estimating the missing year
information associated with publications (and hence the authors' years
of active publication).

Recently, data cleaning on academic social networks has received much
attention. In KDD Cup 2013, the two challenges are the Author-Paper
Identification Challenge and the Author Disambiguation Challenge.
For both challenges, the publishing year information of each paper
is important background knowledge for the design of algorithms.
However, the given data set~\cite{kddcup2013} has a high
\emph{Missing Year Ratio}, $\frac{155784}{2257249}\approx 6.90\%$
(there are in total 2,257,249 papers, and out of which, 155,784 are
missing year papers). This is important motivation for developing
algorithms to recover the missing year attribute of publications, we
called this the Missing Year Estimation (MYE) problem.

The occurrence of the missing data in the bibliographic data can be
caused by a variety of reasons. We believe one reason is that the cited
papers are also included in the dataset, even if the original source
is unavailable. References are sometimes incomplete, leading to
missing and erroneous data. It is also possible that some papers are
recovered from scanned sources which makes it hard to extract all
attributes.

We first propose a simple algorithm that only makes use of the
``direct'' information, such as paper citation/reference
relationships or paper-author relationships. The result of this
simple algorithm is used as a benchmark for comparison. Our goal is
to develop sophisticated algorithms that increase both the coverage
(measured by the percentage of missing year papers recovered) and
accuracy (mean absolute error, or MAE, of the estimated year to the
real year).
The more advanced algorithms we propose and study involve
information propagation rules so that information which is multiple
hops away can also be utilized.
For each algorithm, we propose three versions according to the given
academic social network type: a) Homogenous (only contains paper
citation links), b) Bipartite (only contains paper-author
relations), and, c) Heterogeneous (both paper citation and
paper-author relations). We carry out experiments on the three
public data sets (MSR Libra, DBLP and APS), by applying the K-fold
cross validation method.

Our contributions are: we formulate the problem and introduce a
basic (benchmark) algorithm that can already recover most of the
missing years if both citation and author information are available.
We then systematically developed improved algorithms based on
methods in machine learning. These advanced algorithms further
improve both coverage and accuracy (around $20\%$ in the paper
citation network, $8\%$ in paper author bipartite network and
heterogeneous network), over the benchmark algorithm. In addition,
the coverage achieved by the advanced algorithms well matches the
results derived by the analytical model.

The remaining of the paper is organized as follows; we first
introduce the estimation methodology in section~\ref{Sec:method}, then we
describe the data sets used and the experiment results in
section~\ref{Sec:exp}. In section~\ref{Sec:related}, we discuss the
related works and lastly conclude our work in section~\ref{Sec:conclusion}.

\section{Methodology}\label{Sec:method}
In this section, we first introduce the notations and the three
types of the academic social networks we are dealing with.
For each network type, we propose three corresponding missing year
estimation (MYE) algorithms, with different complexity levels.

\subsection{Notations and three types of the network}
In a general academic social network, there are many types of nodes
and edges. For example, node types can be papers, authors and
publishing venues, etc; and edges can be citations (linking papers
to the papers they cite; authorships (connecting authors to the
papers they have written), and so on.

In the MYE problem, we are mainly interested in two node types:
papers and authors; and two edge types: paper citations and paper
authorships, which induce three academic social networks:
\begin{enumerate}
\item [a)] Paper citation network, denoted by a directed graph
$G_P = (V_P, E_P)$, where $V_P$ is the set of papers and $E_P$ is
the set of citation links. Since citation links have directions,
each citation link can be represented by an ordered paper
pair\footnote{Throughout the paper, we will adopt this special order
of the paper pair for representing the citation links. The reason for this is
that we try to keep this order consistent with the increasing time
line, e.g. a paper with an earlier (left position) publishing time
is cited by a later one (at a right position on the time line).},
i.e., $\forall e = (t, f) \in E_P$, where $t, f \in V_P$, meaning
this citation link is pointing to paper $t$ and originated from
paper $f$.

\item[b)] Paper authorship network, denoted by $G_{AP} = (V_A \cup V_P,
E_{AP})$, where $V_A$ is the set of authors, $V_P$ is the set of
papers and edges in the set $E_{AP}$ connecting authors to their
produced papers (authorship). Hence $G_{AP}$ is a bipartite graph
and we have $\forall e = (a, p) \in E_{AP}$, where $a \in V_A$ and
$p \in V_P$.

\item[c)] Heterogenous network, consisting of both paper citation network and paper authorship
network, denoted by $G = (V_A \cup V_P, E_P \cup E_{AP})$.
\end{enumerate}

Papers are further categorized into two exclusive sets: with known
year information $V_P^K$ and unknown (missing) year information
$V_P^U$. Hence we have $V_P = V_P^K \cup V_P^U$ and $V_P^K \cap
V_P^U = \emptyset$. The remaining notations are listed in
Table~\ref{Tab:Notation}:
\begin{table}[htb]
\centering
\caption{List of Notations}
\begin{tabular}{|c|l|}
\hline
$Y(p),\;\forall\;p \in V_P$ & the real publishing year of paper
$p$, note:\\
& $\forall\;p^U \in V_P^U, Y(p^U)$ is only used for validation purpose.\\
\hline
$T(p),\;\forall\;p \in V_P$ & the set of papers that cite paper $p$,\\
& i.e., $T(p) = \{f| \forall f \in V_P, s.t., (p, f)
\in E_P\}$.\\
\hline
$F(p),\;\forall\;p \in V_P$ &the set of papers that are cited
by paper $p$,\\
& i.e., $F(p) = \{t| \forall t \in V_P, s.t., (t, p) \in E_P\}$.\\
\hline
$\hat{Y}(p^U),\;\forall\;p^U \in V_P^U$ & the estimation result for the missing year paper $p^U$.\\
\hline
$P(a),\;\forall\;a\in V_A$ &the paper set that are written by author $a$.\\
\hline
$A(p),\;\forall\;p\in V_P$ &the author set that have written paper $p$.\\
\hline
$w(p, q), \;\forall p,q\;\in V_P$    & the Consistent-Coauthor-Count between two papers, \\
 & $w(p,q)=w(q,p)=|A(p)\cap A(q)|$\\
\hline
$\Omega(p),\;\forall\;p \in V_P$ & the Consistent-Coauthor-Pair set
of a paper $p \in V_P$,\\
& $\Omega(p) = \{q|q \in V_P\;\textrm{and}\;w(p,q) > 1\}$\\
\hline
$AW\_Min(a),\;AW\_Max(a),$ & the lower and upper bounds of the active publishing\\
$\forall\;a\in V_A$ & time window of author $a$.\\
\hline
$\hat{Y}_{CMin}(p^U), \hat{Y}_{CMax}(p^U),$ & the lower and upper
bounds of the year estimation \\
$\forall\;p^U \in V_P^U$ &  window, derived in the paper citation network $G_P$.\\
\hline
$\hat{Y}_{AMin}(p^U),\;\hat{Y}_{AMax}(p^U)$,
& the lower and upper bounds of the year estimation\\
$\forall\;p^U \in V_P^U$ & window, derived in the paper authorship network $G_{AP}$\\
\hline
$\hat{Y}_{GMin}(p^U),\;\hat{Y}_{GMax}(p^U)$,
& the lower and upper bounds of the year estimation\\
$\forall\;p^U \in V_P^U$ & window, derived in the heterogenous network $G$\\
\hline
\end{tabular}
\label{Tab:Notation}
\end{table}

\subsection{MYE for citation network $G_P$}
\begin{figure}[hbt]
    \centering
    \includegraphics[width=2.4in]{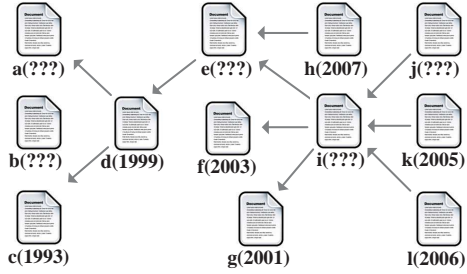}
    \caption{A simple example of a citation network with 12 papers ($a - l$),
            where papers ($a, b, e, i, j$) are $\in V_P^U$ and the remaining
            ($c, d, f, g, h, k, l$) are $\in V_P^K$ .}
    \label{Fig:CWExample}
\end{figure}

We first look at a simple example of the missing year estimation
problem in the paper citation network, shown in
Fig.~\ref{Fig:CWExample}. In this example, there are 12 papers ($a -
l$) and 10 citation edges. 5 papers ($a, b, e, i, j$) have no year
information (i.e. $\in V_P^U$) and the other 7 papers ($c, d, f, g,
h, k, l$) have publishing years (i.e. $\in V_P^K$). Later on, we
will use this example to demonstrate the three MYE algorithms
designed for the citation network $G_P$.

The main idea of estimating the missing years in the citation
network $G_P$ is to make use of paper citing activities, stated as
Assumption~\ref{assumption1}, together with the available
information: a) the year information of those known papers; b) the
citation relationships (edges of the $G_P$).
\begin{assumption}
Normally\footnote{Since the exceptions are rare, we believe that
ignoring such exceptions is reasonable and does not harm our
algorithm design.}, a paper can only cite those papers published
before it, i.e., Eq.~(\ref{assumption1}) is satisfied:
\begin{equation} \label{assumption1}
Y(t) \leq Y(f),\;\;\forall\;\;e = (t, f) \in E_P,\;\;t, f \in V_P.
\end{equation}
\end{assumption}
Assumption~\ref{assumption1} provides the way to determine either a
possible upper bound of the target paper's missing year when it is
cited by a known year paper (i.e., $t\in V_P^U$ and $f\in V_P^K$);
or a possible lower bound of the target paper's missing year when it
cites a known year paper (i.e., $t\in V_P^K$ and $f\in V_P^U$). For
example, using Fig.~\ref{Fig:CWExample}, we can look at paper $a$
(missing year) and $d$ (published in 1999) with a citation link from
$d$ to $a$, we take 1999 as one possible upper bound of $a$'s
publishing year, i.e., $Y(a) \leq 1999$. Similarly, when we look at
paper $d$ and $e$, we get a lower bound of the real publishing year
of $e$, i.e., $1999 \leq Y(e)$.

Following this logic, the missing year estimation task can be
separated into two steps: (1) deriving the possible year estimation
window (two bounds); (2) calculating the missing year value based on
the derived window.

For each step, we propose two methods with different complexity, the
simple (``Sim'') version and the advanced (``Adv'') version. In the
next three subsections, we will introduce the three algorithms
designed for MYE in paper citation network $G_P$. The three
algorithms are different combinations of the two methods in each
step, listed in Table~\ref{Tab:GPCombination}.
\begin{table}[htb]
\centering
\caption{Combination of the two proposed methods in each step, for
the three algorithms for MYE in $G_P$.}
\begin{tabular}{c|c|c}
\hline
Algorithm & Window derivation method & Year value calculation method\\
\hline
$G_P$-SS & Simple & Simple\\
\hline
$G_P$-AS & Advanced & Simple\\
\hline
$G_P$-AA & Advanced & Advanced\\
\hline
\end{tabular}
\label{Tab:GPCombination}
\end{table}
\subsubsection{Algorithm for MYE in $G_P$: $G_P$-SS}
We will first introduce the simple method for each of the two steps,
then we will show how $G_P$-SS works by demonstrating the results on
the example shown in Fig.~\ref{Fig:CWExample}.

{\bf Simple Window Derivation Method:} The simple version of the
window (bounds) derivation method only involves ``one round'' (or in
a ``direct'' manner), which means: (1) spatially, we only consider
those papers that are one-hop to the target missing year paper; (2)
temporally, we only consider immediate (given) information.

Putting together (1) and (2), mathematically, we are deriving the
bounds of the missing year paper $p^U \in V_P^U$ through the subset
of the papers: $F(p^U) \cap V_P^K$ (for the lower bound) and $T(p^U)
\cap V_P^K$ (for the upper bound) as long as they are not empty. For
example, if we look at paper $i$ in Fig.~\ref{Fig:CWExample}, then
only $f$ and $g$ (one-hop away from $i$ and with year information)
are used for deriving the lower bound, while only $k$ and $l$ for
the upper bound. Intuitively, when there are multiple bounds, we
will take the tightest one by applying Eq.~(\ref{eq:y_cmin}) and
(\ref{eq:y_cmax}):
\begin{eqnarray}
\hat{Y}_{CMin}(p^U) &=& \max_{f\;\in\;F(p^U) \cap V_P^K} Y(f),\;
\textrm{if}\;F(p^U) \cap V_P^K\neq\emptyset;\nonumber\\
&=&\;\; -\;\infty,\quad\textrm{otherwise}; \label{eq:y_cmin}\\
\hat{Y}_{CMax}(p^U) &=& \min_{t\;\in\;T(p^U) \cap V_P^K}
Y(t),\;\textrm{if}\;T(p^U) \cap V_P^K\neq\emptyset;\nonumber\\
&=&\;\; +\;\infty,\quad\textrm{otherwise}, \label{eq:y_cmax}
\end{eqnarray}
where $\hat{Y}_{CMin}(p^U)$ denotes the largest possible lower bound
of paper $p^U$ and $\hat{Y}_{CMax}(p^U)$ denotes the smallest
possible upper bound. Here the $-\infty$ and $+\infty$ have no
practical meaning but are used just to represent the non-existent bounds.
In the real implementation, they can be assigned to some pre-defined
constant variables such as ``Default\_Win\_Min'' and
``Default\_Win\_Max''.

Together with the conditions of non-existent bounds, we thus have
four types of possible year estimation windows:
\begin{eqnarray}
&\textrm{Type-1: }& [\hat{Y}_{CMin}(p^U), \hat{Y}_{CMax}(p^U)]; \nonumber\\
&\textrm{Type-2: }& [\;\;\hat{Y}_{CMin}(p^U), \quad+\;\infty\quad);\nonumber\\
&\textrm{Type-3: }& (\quad-\;\infty,\quad \hat{Y}_{CMax}(p^U)\;\;];\nonumber\\
&\textrm{Type-4: }&
(\quad-\;\infty\quad,\quad+\;\infty\quad).\nonumber
\end{eqnarray}
The Type-4 window contains no information for estimation,
hence we define \emph{Uncovered Paper} to be those missing year
papers with a Type-4 estimation window. On the other hand, it is
possible to make a proper estimation on the year value for the
missing year papers with Type-1, Type-2 or Type-3 estimation window.

{\bf Simple Year Value Calculation Method:}
Based on the derived possible year estimation window for each
missing year paper $p^U$, the next step is to make a guess on its
real publishing year. The simple calculation method works in a
straightforward way,
Eqs.~(\ref{eq:CWestType1})-(\ref{eq:CWestType4}):
\begin{eqnarray}
&\textrm{Type-1: }& \hat{Y}(p^U) = \frac{\hat{Y}_{CMin}(p^U) + \hat{Y}_{CMax}(p^U)}{2}, \label{eq:CWestType1}\\
&\textrm{Type-2: }& \hat{Y}(p^U) = \hat{Y}_{CMin}(p^U),\label{eq:CWestType2}\\
&\textrm{Type-3: }& \hat{Y}(p^U) = \hat{Y}_{CMax}(p^U),\label{eq:CWestType3}\\
&\textrm{Type-4: }& \emph{Uncovered}\label{eq:CWestType4}.
\end{eqnarray}
In summary, if both bounds exist (Type-1), we take the average of
the two bounds, Eq.~(\ref{eq:CWestType1}) (assuming that $Y(p^U)$
follows any symmetric discrete distribution centered at the middle
point of the possible estimation window). If only one bound exists
(Type-2 or Type-3), we take the bound value as the calculation
result. Otherwise (Type-4), instead of making any random guess, we
label it as (\emph{Uncovered}), which means, the year of such paper
cannot be estimated properly. Later on, in the performance
evaluation section, we shall consider the uncovered ratio ($=
\frac{\textrm{Total \# Uncovered}}{|V_P^U|}$) of all the proposed
algorithms as one of the performance metrics

Considering the example in Fig.~\ref{Fig:CWExample}, we list both
the intermediate and final estimation results conducted by apply
$G_P$-SS in Table \ref{Tab:GpSS_example}.
\begin{table}[htb]
\centering
\begin{tabular}{|r|c|c|c|c|c|}
 \hline
$p^U$ in Fig.\ref{Fig:CWExample}& $a$   & $b$           & $e$           & $i$           & $j$\\
 \hline
 $F(p^U)$               & $\emptyset$   & $\emptyset$   & $d$           & $e, f, g$     & $i$\\
\hline
 $F(p^U) \cap V_P^K$    & $\emptyset$   & $\emptyset$   & $d$           & $f, g$        & $\emptyset$\\
\hline
 $T(p^U)$               & $d$           & $\emptyset$   & $h, i$        & $j, k, l$     & $\emptyset$\\
\hline
 $T(p^U) \cap V_P^K$    & $d$           & $\emptyset$   & $h$           & $k, l$        & $\emptyset$\\
\hline
 $\hat{Y}_{CMin}(p^U)$  & $-\;\infty$   & $-\;\infty$   & 1999          & 2003          & $-\;\infty$\\
\hline
 $\hat{Y}_{CMax}(p^U)$  & 1999          & $+\;\infty$   & 2007          & 2005          & $+\;\infty$\\
\hline
 $\hat{Y}(p^U)$         & 1999          & \emph{Uncovered}& 2003        & 2004          & \emph{Uncovered}\\
\hline
\end{tabular}
\caption{The intermediate and estimation results obtained through
$G_P$-SS algorithm running on the example of
Fig.~\ref{Fig:CWExample}}
\label{Tab:GpSS_example}
\end{table}

In Table~\ref{Tab:GpSS_example}, the first row lists all the 5
papers belonging to $V_P^U$. The second and third rows list the
paper set cited by each of the 5 papers, where the third row only
contains papers with year information, e.g. for paper $i$, it cites
three papers $F(i) = \{e, f, g\}$ and only two of them have year
information, $F(i) \cap V_P^K = \{f, g\}$. The fourth and fifth rows
list the papers that cite each of the 5 papers, where the fifth row
only contains papers belonging to $V_P^K$. The next two rows are the
two bounds of the possible estimation window by applying
Eqs.~(\ref{eq:y_cmin}) and (\ref{eq:y_cmax}), e.g.,
$\hat{Y}_{CMin}(i) = \max\{Y(f), Y(g)\} = \max\{2003, 2001\} =
2003$. The last row shows the results derived by the simple year
calculation scheme,
Eqs.~(\ref{eq:CWestType1})-(\ref{eq:CWestType4}).

The $G_P$-SS is simple, quick and easy for both implementation and
understanding, but its limitation is also obvious. It has not fully
utilized the available information, which results in a high
uncovered ratio ($= 2/5$ shown in Table~\ref{Tab:GpSS_example}) and
looser bounds. Considering this question, can the information
(derived bounds or estimated results after running $G_P$-SS) of
paper $i$ be useful for its missing year neighbour papers $j$ and
$e$? The answer is yes and the next algorithm is designed to
deal with this.

\subsubsection{Algorithm for MYE in $G_P$: $G_P$-AS}
Comparing to $G_P$-SS, $G_P$-AS applies the same simple version of
year value calculation method,
Eqs.~(\ref{eq:CWestType1})-(\ref{eq:CWestType4}), but an advanced
method for window derivation with information propagations.

A quick way of extending $G_P$-SS is to simply repeat running it. In
this way, the estimated result for a missing year paper (e.g. $i$ in
Fig.~\ref{Fig:CWExample}) in the previous rounds can be used to
derive bounds for its neighbour missing year papers (e.g. $j$ and $e$
in Fig.~\ref{Fig:CWExample}) in the subsequent rounds. However,
since the estimated year result for $i$ can be inaccurate, this kind
of repeating will definitely propagate and even amplify the
inaccuracy.

{\bf Advanced Window Derivation Method:}
Generally in $G_P$, for each citation edge linking two papers, there
can be three possible conditions: (a) both papers have year
information ($\in V_P^K$); or (b) both papers are missing year ($\in
V_P^U$); or (c) one has year information while the other has not.
The limitation of simple window derivation method is that it will only
work under condition (c). By rephrasing Eq.~(\ref{assumption1}) as
Eq.~(\ref{assumption1New}), the advanced window derivation method
relaxes this limitation without inducing any inaccuracy in the
propagation.
\begin{equation}\label{assumption1New}
\hat{Y}_{CMin}(t) \;\leq\; Y(t) \;\leq\; Y(f) \;\leq\;
\hat{Y}_{CMax}(f).
\end{equation}

The rationale behind Eq.~(\ref{assumption1New}) is to extend the
bound transmission rule between two missing year papers: (a) if
$\hat{Y}_{CMin}(t)$ exists, it is also a lower bound of $f$; (b) if
$\hat{Y}_{CMax}(f)$ exists, it is also an upper bound of $t$. The
pseudo code of the advanced window derivation method is included
below.
\begin{algorithm}
\caption{The pseudo code of advanced window derivation method}
\label{Alg:GPAS}
\begin{algorithmic}[1]
\REPEAT
    \STATE UpCnt $\leftarrow 0$;
    \FORALL{$e = (t, f) \in E_P, t,f\;\in V_P$}
        \STATE f\_CMin\_Before$\leftarrow \hat{Y}_{CMin}(f)$;
        \STATE t\_CMax\_Before$\leftarrow \hat{Y}_{CMax}(t)$;
        \IF{$t, f \in V_P^U$}
            \STATE $\hat{Y}_{CMin}(f) \leftarrow \max\{\hat{Y}_{CMin}(f), \hat{Y}_{CMin}(t)\}$;
            \STATE $\hat{Y}_{CMax}(t) \leftarrow \min\{\hat{Y}_{CMax}(t), \hat{Y}_{CMax}(f)\}$;
        \ELSIF{$t \in V_P^K$, $f \in V_P^U$}
            \STATE $\hat{Y}_{CMin}(f) \leftarrow \max\{\hat{Y}_{CMin}(f), Y(t)\}$;
        \ELSIF{$t \in V_P^U$, $f \in V_P^K$}
            \STATE $\hat{Y}_{CMax}(t) \leftarrow \min\{\hat{Y}_{CMax}(t), Y(f)\}$;
        \ENDIF
        \STATE /* Check update counts. */;
        \IF{$\hat{Y}_{CMin}(f) \neq\;\;$f\_CMin\_Before}
            \STATE UpCnt $\leftarrow$ UpCnt $ + 1$;
        \ENDIF
        \IF{$\hat{Y}_{CMax}(t) \neq\;\;$t\_CMax\_Before}
            \STATE UpCnt $\leftarrow$ UpCnt $ + 1$;
        \ENDIF
    \ENDFOR
\UNTIL{UpCnt$\;=\;0$; /* When no update happens, loop ends. */}
\end{algorithmic}
\end{algorithm}

In Algorithm~\ref{Alg:GPAS}, we first initialize a local variable
``UpCnt'' which records the total number of bound updates in each
loop (Line 2). Lines 3-21 are steps in a loop of processing each
citation link of $G_P$, where Lines 9-13 are the same as the simple
window derivation method, Eq.~(\ref{eq:y_cmin}) and
Eq.~(\ref{eq:y_cmax}), while Lines 6-8 are the essential part that
differs from the simple version (also the implementation of the two
bound transmission rules of Eq.~(\ref{assumption1New})).

In Table~\ref{Tab:GpAS_example}, we list both the intermediate and
estimation results of applying $G_P$-AS on the example of
Fig.~\ref{Fig:CWExample}.
\begin{table}[htb]
\centering
\begin{tabular}{|c|c|c|c|c|}
 \hline
  $p^U$ in Fig.\ref{Fig:CWExample} & Round 1      & Round 2       & Round 3 & $\hat{Y}(p^U)$\\
 \hline
 $a$      & $(-\infty, 1999)$     & $(-\infty, 1999)$     & $(-\infty, 1999)$ & 1999\\
\hline
 $b$      & $(-\infty, +\infty)$  & $(-\infty, +\infty)$  &$(-\infty, +\infty)$ & \emph{NotCovered}\\
\hline
 $e$      & $(1999, 2007)$        & $(1999, 2005)$        & $(1999, 2005)$ & 2002\\
\hline
 $i$      & $(2003, 2005)$        & $(2003, 2005)$        & $(2003, 2005)$ & 2004\\
\hline
 $j$      & $(-\infty, +\infty)$  & $(2003, +\infty)$     & $(2003, +\infty)$ & 2003 \\
\hline
 & UpCnt = 5                     & UpCnt =2                     &UpCnt =0& \\
\hline
\end{tabular}
\caption{The intermediate and estimation results of applying
$G_P$-AS on the example shown in Fig.~\ref{Fig:CWExample}}
\label{Tab:GpAS_example}
\end{table}

From Table~\ref{Tab:GpAS_example}, we can see that the advanced
window estimation takes two rounds (no updates happen in round 3)
and the last column is the year estimation results by applying the
simple year value calculation method based on the derived bounds.
Comparing to Table~\ref{Tab:GpSS_example}, the improvement is
obvious even for this simple example: (1) paper $j$ is no longer
labeled as (\emph{Uncovered}, hence, the uncovered ratio decreases
to 1/5; (2) paper $e$ gets a tighter possible estimation window.

So far, we are doing our best to deal with the possible window
derivation problem (apparently, paper $b$ in
Fig.~\ref{Fig:CWExample} has no chance of getting a good estimate, and
we will discuss the relationship between the uncovered ratio and the
structure of the given citation graph $G_P$ mathematically in
Section~\ref{Sec:exp}). In the next algorithm, we investigate how
the year value calculation method can be further improved.

\subsubsection{Algorithm for MYE in $G_P$: $G_P$-AA}
Given the derived estimation window $[\hat{Y}_{CMin}(p^U),
\hat{Y}_{CMax}(p^U)]$ for a missing year paper $p^U$, recall
Eqs.~(\ref{eq:CWestType1})-(\ref{eq:CWestType4})(how simple year
value calculation method works): (1) if both bounds exist (Type-1),
the calculation result is the mean of the two bounds; or (2) if only
one bound exists (Type-2 or Type-3), the calculation result equals
to the value of the existing bound; or (3) if neither bound exists,
then the paper is labeled as \emph{Uncovered}, representing no
proper estimation result.

The year estimation results for cases (1) and (2) affect the
accuracy metrics, such as Mean Absolute Error (MAE), while case (3)
only affects the uncovered ratio, irrelevant to other metrics. For
case (1), it is rational to take the average of the two bounds,
since the citing-to activity and cited-by activity can be considered
symmetric. However for case (2), more investigation is needed. The
physical interpretation of case (2) is based on the assumption that
the missing year paper has the same publishing time as the earliest
paper that cites it (the upper bound exists), or the latest paper
cited by it (the lower bound exists). In reality, this seldom
happens. The best guess for a (Type-2 or Type-3) window case may be
correlated to the bound value, not just a fixed distance to the
bound (e.g. the simple calculation method takes a fixed zero
distance). Therefore, the solution for this problem is to find a
proper function $\hat{y}(p^U) = d(WinType(p^U), BoundVal(p^U))$ to
calculate $\hat{y}(p^U)$ for each missing year paper $p^U$, based on
its derived estimation window type, denoted by $WinType(p^U)$ (which
takes value of either Type-2 or Type-3), and the value of bound,
denoted by $BoundVal(p^U)$.

To achieve this, we need a separate data set, denoted by
$\mathcal{T}$, containing a series of 3-tuple data $t = \{y_{t},
WinType_{t}, BoundVal_{t}\} \in \mathcal{T}$ for training purpose.
Each 3-tuple data corresponds to a missing year paper $t$ in this
training set, where $y_t$ is the validated real publishing year,
$WinType_t$ is the derived estimation window type and $BoundVal_t$
is the bound value. If we denote $\mathcal{T}_{p^U}$ as the subset
of $\mathcal{T}$ with respect to $p^U$ and $\mathcal{T}_{p^U} =
\{t|t\in{\mathcal{T}}, WinType_{t} = WinType(p^U), BoundVal_{t} =
BoundVal(p^U)\}$, then we get the following form for $d(\cdot)$
corresponding with $\mathcal{T}$:
\begin{equation}\label{eq:d_func}
\hat{y}(p^U) = d_{\mathcal{T}}\big(WinType(p^U),
BoundVal(p^U)\big) =
\frac{\sum_{t\in\mathcal{T}_{p^U}}y_t}{|\mathcal{T}_{p^U}|},
\end{equation}
where $|\mathcal{T}_{p^U}|$ is the element count of the set
$\mathcal{T}_{p^U}$.

The idea of Eq.~(\ref{eq:d_func}) is to take the expectation of the
real publishing years of those papers having the same window type
and bound value as $P^U$ in the training set $\mathcal{T}$. However
it is not trivial to find a proper training set satisfying: (1) a
citation graph with similar property and structure to the given
$G_P$; (2) the $BoundVal$ of this training set covers a wider range
than that of $BoundVal(p^U), \forall p^U \in V_P^U$.

{\bf Advanced Year Value Calculation Method:}
We first propose a way to find a suitable training set $\mathcal{T}$
which can satisfy both (1) and (2) mentioned above. After that, the
estimation results can be calculated through Eq.~(\ref{eq:d_func}).

One of the most suitable training sets is just inside the given
citation network $G_P$. In fact, each paper with known year
($\forall p^K \in V_P^K$) can also be used to derive a possible
estimation window (by pretending itself to be a missing year paper).
Consider the example in Fig.~\ref{Fig:CWExample}, for paper
$d(1999)$, the simple window derivation method generates $[1993,
+\infty)$. Since this is independent of deriving windows for missing
year papers, these two procedures can be merged together to save the
running time. The modified advanced window derivation method for
$G_P$-AA is shown in Algorithm~\ref{Alg:GPAA}.
\begin{algorithm}
\caption{The modified advanced window derivation method for
$G_P$-AA} \label{Alg:GPAA}
\begin{algorithmic}[1]
\REPEAT
    \STATE UpCnt $\leftarrow 0$;
    \FORALL{$e = (t, f) \in E_P, t,f\;\in V_P$}
        \STATE f\_CMin\_Before$\leftarrow \hat{Y}_{CMin}(f)$;
        \STATE t\_CMax\_Before$\leftarrow \hat{Y}_{CMax}(t)$;
        \IF{$t, f \in V_P^U$}
            \STATE $\hat{Y}_{CMin}(f) \leftarrow \max\{\hat{Y}_{CMin}(f), \hat{Y}_{CMin}(t)\}$;
            \STATE $\hat{Y}_{CMax}(t) \leftarrow \min\{\hat{Y}_{CMax}(t), \hat{Y}_{CMax}(f)\}$;
        \ELSIF{$t \in V_P^K$, $f \in V_P^U$}
            \STATE $\hat{Y}_{CMin}(f) \leftarrow \max\{\hat{Y}_{CMin}(f), Y(t)\}$;
            \STATE $\hat{Y}_{CMax}(t) \leftarrow \min\{\hat{Y}_{CMax}(t), \hat{Y}_{CMax}(f)\}$;
            /* for training set $\mathcal{T}$. */
        \ELSIF{$t \in V_P^U$, $f \in V_P^K$}
            \STATE $\hat{Y}_{CMin}(f) \leftarrow \max\{\hat{Y}_{CMin}(f), \hat{Y}_{CMin}(t)\}$;
            /* for training set $\mathcal{T}$. */
            \STATE $\hat{Y}_{CMax}(t) \leftarrow \min\{\hat{Y}_{CMax}(t), Y(f)\}$;
        \ELSE[$t, f \in V_P^K$]
            \STATE $\hat{Y}_{CMin}(f) \leftarrow \max\{\hat{Y}_{CMin}(f), Y(t)\}$;
            /* for training set $\mathcal{T}$. */
            \STATE $\hat{Y}_{CMax}(t) \leftarrow \min\{\hat{Y}_{CMax}(t), Y(f)\}$;
            /* for training set $\mathcal{T}$. */
        \ENDIF
        \STATE /* Check update counts. */;
        \IF{$\hat{Y}_{CMin}(f) \neq\;\;$f\_CMin\_Before}
            \STATE UpCnt $\leftarrow$ UpCnt $ + 1$;
        \ENDIF
        \IF{$\hat{Y}_{CMax}(t) \neq\;\;$t\_CMax\_Before}
            \STATE UpCnt $\leftarrow$ UpCnt $ + 1$;
        \ENDIF
    \ENDFOR
\UNTIL{UpCnt$\;=\;0$; /* When no update happens, loop ends. */}
\end{algorithmic}
\end{algorithm}

Comparing to Algorithm~\ref{Alg:GPAS}, the pseudo code in
Algorithm~\ref{Alg:GPAA} has added 4 lines (Lines 11, 13, 16 and 17)
for preparing the training set. These four lines are still
satisfying Eq.~(\ref{assumption1New}) for avoiding inducing
inaccuracy, but the information is propagated towards papers in set
$V^K_P$. Table~\ref{Tab:GPAA_Example} lists the intermediate and
final results of the example training set $\mathcal{T}$ in
Fig.~\ref{Fig:CWExample}.

\begin{table}[htb]
\centering
\begin{tabular}{|c|c|c|c|c|c|c|}
 \hline
  $p^K$ in Fig.\ref{Fig:CWExample} & Round 1      & Round 2       & Round 3 & Round 4 &$WinType$\\
 \hline
 $c(1993)$   & $(-\infty, 1999)$       & $(-\infty, 1999)$     & $(-\infty, 1999)$& $(-\infty, 1999)$ & Type-3\\
\hline
 $d(1999)$   & $(1993, +\infty)$       & $(1993, 2007)$        &$(1993, 2005)$&$(1993, 2005)$ & Type-1 \\
\hline
 $f(2003)$   & $(-\infty, +\infty)$    & $(-\infty, 2005)$     & $(-\infty, 2005)$& $(-\infty, 2005)$& Type-3\\
\hline
 $g(2001)$   & $(-\infty, +\infty)$    & $(-\infty, 2005)$     & $(-\infty, 2005)$& $(-\infty, 2005)$& Type-3\\
\hline
 $h(2007)$   & $(-\infty, +\infty)$    & $(1999, +\infty)$     & $(1999, +\infty)$& $(1999, +\infty)$& Type-2\\
\hline
 $k(2005)$   & $(-\infty, +\infty)$    & $(2003, +\infty)$     & $(2003, +\infty)$&$(2003, +\infty)$& Type-2\\
\hline
 $l(2006)$   & $(-\infty, +\infty)$    & $(2003, +\infty)$     & $(2003, +\infty)$ & $(2003, +\infty)$& Type-2\\
\hline
       & UpCnt = 2                & UpCnt = 6                     &UpCnt =1& UpCnt =0& \\
\hline
\end{tabular}
\caption{The intermediate and final results of the example training
set $\mathcal{T}$ in Fig.~\ref{Fig:CWExample}.}
\label{Tab:GPAA_Example}
\begin{tabular}{|c|c|c|}
 \hline
  $p^U$ in Fig.\ref{Fig:CWExample} & $a$& $j$ \\
\hline
 Derived Window & $(-\infty, 1999)$ & $(2003, +\infty)$\\
\hline
$WinType/BoundVal$ & Type-3/1999 & Type-2/2003\\
\hline
$\hat{Y}(p^U)$ by $G_P$-AS & 1999& 2003\\
\hline
$\mathcal{T}_{p^U}$ &c(1993, Type-3, 1999)  & k(2005, Type-2, 2003)\\
                    &                       & l(2006, Type-2,
                    2003)\\
\hline
$\hat{Y}(p^U)$ by $G_P$-AA & 1993 & 2006 (2005.5)\\
\hline
\end{tabular}
\caption{Comparison on the estimation results on papers $a$ and $j$
of the example in Fig.~\ref{Fig:CWExample} by $G_P$-AS versus
$G_P$-AA.} \label{Tab:GPAA_Example2}
\end{table}

Recall Table~\ref{Tab:GpAS_example}, we notice that the estimation
results of paper $a$ and paper $j$ will be affected by the advanced
year value calculation method, according to the derived training set
in Table~\ref{Tab:GPAA_Example} and Eq.~(\ref{eq:d_func}). The
comparison on the estimation results between $G_P$-AS and $G_P$-AA
is listed in Table~\ref{Tab:GPAA_Example2}.

So far, we are only illustrating how the three algorithms work and
how different the estimation results appear. In the experiment
section (Section~\ref{Sec:exp}), we will see their performance
evaluated on the real data sets.

\newpage
\subsection{MYE for paper authorship network $G_{AP}$}
\begin{figure}[hbt]
    \centering
    \includegraphics[width=2.6in]{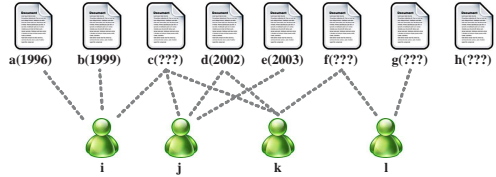}
    \caption{An example of a paper authorship network with 8 papers ($a -
    h$) and 4 authors ($i - l)$, where papers ($a, b, d, e$) are $\in V_P^K$ and ($c, f, g, h$) are $\in
    V_P^U$.}
    \label{Fig:AWExample}
\end{figure}

In this section, we move to the paper-author bipartite graph
$G_{AP}$. An artificially created example of MYE problem in $G_{AP}$
is shown in Fig.~\ref{Fig:AWExample}. In this example, there are 8
papers ($a - h$) and 4 authors ($i - l)$, where papers $a, b, d, e$
have year information ($\in V_P^K$) while $c, f, g, h$ are missing
year ($\in V_P^U$).

For $G_{AP}$, we will also introduce three algorithms, namely
$G_{AP}$-Ba, $G_{AP}$-Iter and $G_{AP}$-AdvIter, in order of increasing
complexity.

\subsubsection{Algorithm for MYE in $G_{AP}$:
$G_{AP}$-Ba and $G_{AP}$-Iter}
$G_{AP}$-Ba is the basic algorithm and $G_{AP}$-Iter is simply
repeating $G_{AP}$-Ba until convergence, thus we introduce them
together. The basic algorithm, $G_{AP}$-Ba, consists of three steps:
\begin{enumerate}
\item[i)] Derive author's active publishing window.

For each author, based on the graph topology and paper year
information, we can derive an active paper publishing window.
Eqs.~(\ref{eq:awmin}) and (\ref{eq:awmax}) give the definition of
the two bounds of this window:
\begin{eqnarray}
AW\_Min(a) = \min_{p\in P(a) \cap V_P^K} Y(p),\label{eq:awmin}\\
AW\_Max(a) = \max_{p\in P(a) \cap V_P^K} Y(p),\label{eq:awmax}
\end{eqnarray}
where $P(a),\forall\;a \;\in V_A$ is the paper set written by author
$a$. It is possible that $P(a) \cap V_P^K = \emptyset$, and we
consider it as a non-existent bound. According to the above
definition, the two bounds are either co-existent or non-existent.\\

\item[ii)] Derive the paper's possible year estimation window.

Based on the derived author active window, we can further define the
paper possible year window:
\begin{eqnarray}
\hat{Y}_{AMin}(p^U) = \min\{\max_{a\in A(p^U)} AW\_min(a), \min_{a\in A(p^U)} AW\_max(a)\},\label{eq:acwmin}\\
\hat{Y}_{AMax}(p^U) = \max\{\max_{a\in A(p^U)} AW\_min(a),
\min_{a\in A(p^U)} AW\_max(a)\},\label{eq:acwmax}
\end{eqnarray}
where $A(p^U), \forall p^U \in V_P^U$ is the author set of paper
$p^U$.\\

In most cases, $\hat{Y}_{AMin}(p^U) = \max_{a\in A(p^U)} AW\_min(a)$
and $\hat{Y}_{AMax}(p^U) = \min_{a\in A(p^U)} AW\_max(a)$. However,
in case of the condition that authors' active windows have no
intersection (this is possible because the author active window dose
not take the missing year papers into account), we rewrite them as Eqs.~(\ref{eq:acwmin})-(\ref{eq:acwmax}). For example, we look at
paper $c$ in Fig.~\ref{Fig:AWExample}. The author set of paper $c$
is $A(c) = \{i(1996,1999), j(2002, 2003), k(2002, 2002)\}$ with the
author active windows inside parentheses. Then by definition we get
$\max_{a\in A(c)} AW\_Min(a) = \max\{1996, 2002, 2002\} = 2002$,
while $\min_{a\in A(c)} AW\_Max(a) = \min\{1999, 2003, 2002\} =
1999$. Therefore, according to
Eqs.~(\ref{eq:acwmin})-(\ref{eq:acwmax}), we can derive the possible
year estimation window of paper $c: [1999, 2002]$.\\

\item[iii)] Calculate year value.

In this algorithm, we apply the simple year value calculation
method, the same one as in the $G_P$-SS algorithm. There is only a
small difference in that in $G_P$-SS, there are four types of the year
estimation window, whereas in $G_{AP}$, there are only two possible
types, both bounds exist (Type-1) or neither exists (Type-4).
Therefore, the estimated year value is either
$\frac{\hat{Y}_{AMin}(p^U) + \hat{Y}_{AMax}(p^U)}{2}$ or labeled as
\emph{Uncovered}.
\end{enumerate}

Note the rationale of the design of the basic algorithm is based on
an observation that most authors are continuously active in
publishing papers. Hence, the publishing years of his/her papers are
usually within a continuous window. If we obtain the windows of all
the coauthors of a missing year paper, the intersection of these
windows will be an interval that most probably contains the real
publishing year.

\begin{algorithm}
\caption{The pseudo code of $G_{AP}$-Iter} \label{Alg:AWProp}
\begin{algorithmic}[1]
 \REPEAT
    \FORALL{$e = (a, p) \in E_{AP}, a \in V_A, p \in V_P$}
        \IF{$p \in V_P^K$}
            \STATE $AW\_Min(a) \leftarrow \min\{Y(p), AW\_Min(a)\}$
            \STATE $AW\_Max(a) \leftarrow \max\{Y(p), AW\_Max(a)\}$
        \ELSIF[$p \in V_P^U$]{$\hat{Y}(p)$ exists}
            \STATE $AW\_Min(a) \leftarrow \min\{\hat{Y}(p), AW\_Min(a)\}$
            \STATE $AW\_Max(a) \leftarrow \max\{\hat{Y}(p), AW\_Max(a)\}$
        \ENDIF
    \ENDFOR
    \FORALL{$p^U \in V_P^U$}
        \FORALL{$a \in A(p^U)$}
            \STATE $maxMin \leftarrow \max\{AW\_Min(a), maxMin\}$
            \STATE $minMax \leftarrow \min\{AW\_Max(a), minMax\}$
        \ENDFOR
        \STATE $\hat{Y}_{AMin}(p^U) \leftarrow \min\{maxMin, minMax\}$
        \STATE $\hat{Y}_{AMax}(p^U) \leftarrow \max\{maxMin, minMax\}$
        \STATE $\hat{Y}(p^U) \leftarrow \frac{\hat{Y}_{AMin}(p^U) + \hat{Y}_{AMax}(p^U)}{2}$
    \ENDFOR
\UNTIL{No update happens}
\end{algorithmic}
\end{algorithm}
The pseudo code for $G_{AP}$-Iter (including $G_{AP}$-Ba) is shown
in Algorithm~\ref{Alg:AWProp}. Lines 2-19 are the steps of
$G_{AP}$-Ba, $G_{AP}$-Iter is simply repeating $G_{AP}$-Ba (Line
1). The estimation results in the previous rounds affect the
subsequent rounds because each author's active publishing window
will be re-calculated according to all the paper year information
(given or estimated in the last round, Lines 7-8). Lines 13-17 are
the implementation of Eqs.~(\ref{eq:acwmin}) and (\ref{eq:acwmax}).
The intermediate and final estimation results of running
$G_{AP}$-Iter on the example in Fig.~\ref{Fig:AWExample} is listed
in Table~\ref{Tab:AWPropExample}.

\begin{table}[htb]
\centering
\begin{tabular}{|c|c|c|c|c|}
 \hline
 Node   & Type    & Round 1           & Round 2       & Round 3\\
 \hline
 $i$    & Author  & $(1996, 1999)$    & $(1996, 2001)$     & $(1996, 2001)$\\
\hline
 $j$    & Author  & $(2002, 2003)$    & $(2001, 2003)$     &$(2001, 2003)$\\
\hline
 $k$    & Author  & $(2002, 2002)$    & $(2001, 2002)$     & $(2001, 2002)$\\
\hline
 $l$    & Author  & $(-\infty, +\infty)$ & $(2002, 2002)$ & $(2002, 2002)$\\
\hline \hline
 $c$    & Paper  & $(1999, 2002)$    & $(2001, 2001)$     & $(2001, 2001)$\\
 $\hat{Y}(c)$ &  & 2001 (2000.5)     &  2001 & 2001\\
\hline
 $f$    & Paper  & $(2002, 2002)$    & $(2002, 2002)$     & $(2002, 2002)$\\
 $\hat{Y}(f)$ &  & 2002     &  2002 & 2002\\
\hline
 $g$    & Paper  & $(-\infty, +\infty)$    & $(2002, 2002)$     & $(2002, 2002)$\\
 $\hat{Y}(g)$ &  & \emph{Uncovered}     &  2002 & 2002\\
\hline
 $h$    & Paper  & $(-\infty, +\infty)$    & $(-\infty, +\infty)$  & $(-\infty, +\infty)$\\
 $\hat{Y}(h)$ &  & \emph{Uncovered}     &  \emph{Uncovered} & \emph{Uncovered}\\
\hline
\end{tabular}
\caption{The intermediate and final estimation results obtained by
running $G_{AP}$-Ba and $G_{AP}$-Iter on the example shown in
Fig.~\ref{Fig:AWExample}} \label{Tab:AWPropExample}
\end{table}

In Table~\ref{Tab:AWPropExample}, the $G_{AP}$-Iter repeats 3 rounds
until convergence. We show the intermediate results of the author
active windows for author (nodes $i, j, k , l$), the possible paper
publishing windows for missing year papers (nodes $c, f, g, h$), and
their estimation results ($\hat{Y}(p^U), p^U \in \{c, f, g, h\}$) in
each round. The column labeled as ``Round 1'' shows the results
generated by algorithm $G_{AP}$-Ba. Comparing to $G_{AP}$-Ba,
$G_{AP}$-Iter helps to share information through the co-author
relationships, like author $l$ in Table~\ref{Tab:AWPropExample}.
Therefore, $G_{AP}$-Iter obtains a lower uncovered ratio (1/4) than
$G_{AP}$-Ba (2/4).

We need to note that $G_{AP}$-Iter may add inaccuracy during the
information propagation, i.e. the estimation results in the
previous rounds affect the derivation of both the author active
windows and estimation results in the subsequent rounds. For
example, $\hat{Y}(c)$ after Round 1 is 2001. In Round 2, the active
windows of all the coauthors of paper $c$, $P(c) = \{i, j, k\}$ are
updated, and hence the related paper year estimation windows get
updated also. Although $G_{AP}$-Iter helps to decrease the uncovered
ratio, it may not improve an estimation accuracy like MAE (under
certain situations, $G_{AP}$-Iter can be even worse than $G_{AP}$-Ba).

In order to compensate for the weakness of $G_{AP}$-Iter so that both
the uncovered ratio and estimation accuracies are improved we propose
the $G_{AP}$-AdvIter, which has an advanced iteration procedure to
reduce the propagation of inaccurate information.

\subsubsection{Algorithm for MYE in $G_{AP}$: $G_{AP}$-AdvIter}
According to the previous discussion, the key point of improving the
estimation accuracy in $G_{AP}$ is to propagate as much ``good''
information as possible. Hence, we propose a heuristic algorithm,
$G_{AP}$-AdvIter to achieve this. Here are some definitions:
\begin{enumerate}
\item[1.] Consistent-Coauthor-Count between two papers: the number
of common coauthors of the two papers. We denote it by function
$w(\cdot)$. Given any two papers, we can calculate their
Consistent-Coauthor-Count by the following expression:
\begin{equation}
\forall\;p,\;q\;\in\;V_P,\; w(p, q) = w(q, p) = |A(p) \cap A(q)|,
\end{equation}
where $w(\cdot)$ is a non-negative integer and equals to zero only
when the two papers have no common coauthors.

\item[2.] $w$-Consistent-Coauthor-Pair relationship:
if any two papers, $\forall p,q\in V_P$, satisfy: $w(p, q) = w(q, p)
> 1$, then we call them a $w$-Consistent-Coauthor-Pair.

\item[3.] Consistent-Coauthor-Pair set of a paper $p\in V_P$, denoted
by $\Omega(p)$:
\begin{equation}
\Omega(p) = \{q| q \in V_P\;\textrm{and}\;w(p,q) > 1 \}
\end{equation}
\end{enumerate}
We give some illustrations of these definitions using the example in
Fig.~\ref{Fig:AWExample}: $w(a,g) = |\emptyset| = 0$ and $w(c, d) =
|\{j, k\}| = 2$, thus, paper $c, d$ have the
2-Consistent-Coauthor-Pair relationship. Not including this, there are no
more Consistent-Coauthor-Pairs in Fig.~\ref{Fig:AWExample}.
Therefore, we obtain $\Omega(c) = \{d\}$, $\Omega(d) = \{c\}$ and
$\Omega(p) = \emptyset, \forall p \in \{a,b,e,f,g,h\}$.

It is a reasonable assumption that if more authors work together and
publish papers, it is more probable that these papers are published
within a small time window. For example, students who worked together
with their supervisors/group members and published certain papers
during their Master/PhD study. Note this is only a sufficient
condition, the reverse may not be true.

The above assumption implies that if two papers have a $w$-Consistent-Coauthor-Pair relationship, then there is a high
probability that their publishing years are close. In addition, this
probability is positively correlated to the value of $w$. We
conjecture that the estimated year values from utilizing the
$w$-Consistent-Coauthor-Pair relationship must be ``better''
information for propagation.

The pseudo code of $G_{AP}$-AdvIter is listed in Algorithm
\ref{Alg:AWLearn}, which shows how we make use of the more reliable
information for propagation.
\begin{algorithm}
\caption{The pseudo code of $G_{AP}$-AdvIter} \label{Alg:AWLearn}
\begin{algorithmic}[1]
 \FORALL{$p^U \in V_P^U$}
    \STATE Derive the Consistent-Coauthor-Pair set, $\Omega(p^U)$.
 \ENDFOR
 \REPEAT
    \FORALL{$e = (a, p) \in E_{AP}, a\;\in V_A, p \in V_P$}
        \IF{$p \in V_P^K$}
            \STATE $AW\_Min(a) \leftarrow \min\{Y(p), AW\_Min(a)\}$
            \STATE $AW\_Max(a) \leftarrow \max\{Y(p), AW\_Max(a)\}$
        \ELSIF[$p \in V_P^U$]{$\hat{Y}(p)$ exists}
            \STATE $AW\_Min(a) \leftarrow \min\{\hat{Y}(p), AW\_Min(a)\}$
            \STATE $AW\_Max(a) \leftarrow \max\{\hat{Y}(p), AW\_Max(a)\}$
        \ENDIF
    \ENDFOR
    \FORALL{$p^U \in V_P^U$}
        \IF[for AdvIter]{$\Omega(p^U) \cap V_P^K \neq \emptyset$}
            \STATE $\hat{Y}(p^U) \leftarrow W(p^U, \gamma)$
        \ELSE
            \FORALL{$a \in A(p^U)$}
                \STATE $maxMin \leftarrow \max\{AW\_Min(a), maxMin\}$
                \STATE $minMax \leftarrow \min\{AW\_Max(a), minMax\}$
            \ENDFOR
            \STATE $\hat{Y}_{AMin}(p^U) \leftarrow \min\{maxMin, minMax\}$
            \STATE $\hat{Y}_{AMax}(p^U) \leftarrow \max\{maxMin, minMax\}$
            \STATE $\hat{Y}(p^U) \leftarrow \frac{\hat{Y}_{AMin}(p^U) + \hat{Y}_{AMax}(p^U)}{2}$
        \ENDIF
    \ENDFOR
\UNTIL{No update happens}
\end{algorithmic}
\end{algorithm}

Comparing to Algorithm~\ref{Alg:AWProp}, we notice that Algorithm
\ref{Alg:AWLearn} only added Lines 1-3 and Lines 15-16. Lines 1-3
is the process used to find $\Omega(p^U)$ for each missing year paper,
this is done during initialization. Lines 15-16 show that we
give higher priority to estimating year values if the
$w$-Consistent-Coauthor-Pair relationship can help, than the basic
procedure (Lines 17-25). The expression of the function $W$ is in
Eq.~(\ref{eq:wa}):
\begin{equation}
W(p^U,\gamma) = \frac{\sum_{q\in \Omega(p^U)\cap V_P^K}\;\;w(p^U,
q)^\gamma \times Y(q)}{\sum_{q\in \Omega(p^U)\cap V_P^K} w(p^U,
q)^\gamma}, \quad\textrm{if}\;\Omega(p^U)\cap V_P^K \neq
\emptyset\label{eq:wa}
\end{equation}
The meaning of Eq.~(\ref{eq:wa}) is to take a $\gamma$-weighted
average on the given year information of those papers in the set
$\Omega(p^U)\cap V_P^K$. For example, if $\Omega(p^U)\cap V_P^K =
\{q, r\}, w(p^U, q) = 2, w(p^U, r) = 3, Y(q) = 2000, Y(r) = 2002$,
then $W(p^U,\gamma) = \frac{2^\gamma \times 2000 + 3^\gamma \times
2002}{2^\gamma + 3^\gamma}$. Here parameter $\gamma$ is used to tune
the importance we put on the values of $w$, e.g., if we set $\gamma
= 0$, it implies that no weight is considered and the result is
simply the average. When $\gamma = 1$, it is a normal weighted
average calculation; while when $\gamma\rightarrow\infty$, it leads to
the special case where only the papers in the set $\Omega(p^U)\cap
V_P^K$ with the largest $w$ are involved in the calculation. In
addition, since it is meaningless for function $W$ if
$\Omega(p^U)\cap V_P^K = \emptyset$, we need to check
beforehand (Line 15).
\begin{table}[htb]
\centering
\begin{tabular}{|c|c|c|c|c|}
 \hline
 Node   & Type    & Round 1           & Round 2       & Round 3\\
 \hline
 $i$    & Author  & $(1996, 1999)$    & $(1996, 2002)$     & $(1996, 2002)$\\
\hline
 $j$    & Author  & $(2002, 2003)$    & $(2002, 2003)$     &$(2002, 2003)$\\
\hline
 $k$    & Author  & $(2002, 2002)$    & $(2002, 2002)$     & $(2002, 2002)$\\
\hline
 $l$    & Author  & $(-\infty, +\infty)$ & $(2002, 2002)$ & $(2002, 2002)$\\
\hline \hline
 $c$    & Paper  & $(2002, 1999)$    & $(2001, 2001)$     & $(2001, 2001)$\\
 $\hat{Y}(c)$& $W(c, 0)$    & 2002    &  2002 & 2002\\
\hline
 $f$    & Paper  & $(2002, 2002)$    & $(2002, 2002)$     & $(2002, 2002)$\\
 $\hat{Y}(f)$ &  & 2002     &  2002 & 2002\\
\hline
 $g$    & Paper  & $(-\infty, +\infty)$    & $(2002, 2002)$     & $(2002, 2002)$\\
 $\hat{Y}(g)$ &  & \emph{Uncovered}     &  2002 & 2002\\
\hline
 $h$    & Paper  & $(-\infty, +\infty)$    & $(-\infty, +\infty)$  & $(-\infty, +\infty)$\\
 $\hat{Y}(h)$ &  & \emph{Uncovered}     &  \emph{Uncovered} & \emph{Uncovered}\\
\hline
\end{tabular}
\caption{The intermediate and final estimation results obtained by
running $G_{AP}$-AdvIter on the example shown in
Fig.~\ref{Fig:AWExample}} \label{Tab:AWLearnExample}
\end{table}

In Table~\ref{Tab:AWLearnExample}, we list the intermediate and
final estimation results obtained by running $G_{AP}$-AdvIter on the
example shown in Fig.~\ref{Fig:AWExample}. As analyzed previously,
$\Omega(c) = \{d\}$, $\Omega(d) = \{c\}$ and $\Omega(p) = \emptyset,
\forall p \in \{a,b,e,f,g,h\}$, hence only $\hat{Y}(c) = Y(d) =
2002$ is affected by $G_{AP}$-AdvIter and also the related author
active windows: $i:(1996, 2002)$, $j:(2002, 2003)$ and $k:(2002,
2002)$.

\subsection{MYE for heterogenous network $G$}
For a heterogeneous network, $G = (G_P \cup G_{AP})$, which consists
of both $G_P$ and $G_{AP}$, we make use of the proposed methods and
the results discussed in the previous two sections. Since for both $G_P$
and $G_{AP}$, we proposed three algorithms of different complexity,
there can be altogether 9 different combinations. With careful
consideration, we pick out 3 typical combinations as MYE algorithms
for $G$:
\begin{enumerate}
\item[1)] $G$-SSBa: combination of $G_P$-SS and $G_{AP}$-Ba
\item[2)] $G$-ASIter: combination of $G_P$-AS and $G_{AP}$-Iter
\item[3)] $G$-AdvIter: combination $G_P$-AA and $G_{AP}$-AdvIter
\end{enumerate}

In fact, selecting the ``combination'' is not trivial, this
will be explained in more detail next. The common part of the two algorithms
consists of these two steps: (a) derivation of possible year
estimation window and (b) calculate the estimated year value based
on the derived window.

No matter which combined algorithm for $G$ is applied, for each
missing year paper, two possible year estimation windows will be
derived, one by the $G_P$ part $[\hat{Y}_{CMin}(p^U),
\hat{Y}_{CMax}(p^U)]$, and the other by the $G_{AP}$ part
$[\hat{Y}_{AMin}(p^U), \hat{Y}_{AMax}(p^U)]$, due to the
independency of the two procedures.

Considering the four types of the derived estimation window from
$G_P$ and two types from $G_{AP}$, each missing year paper can end
with the following four cases of which case (d) is most likely:
\begin{enumerate}
\item[(a)] $(\hat{Y}_{CMin}(p^U),
\hat{Y}_{CMax}(p^U)) = (\hat{Y}_{AMin}(p^U), \hat{Y}_{AMax}(p^U)) =
(-\infty, +\infty)$, then it can only lead to the \emph{Uncovered}
estimation result;
\item[(b)] $(\hat{Y}_{CMin}(p^U),
\hat{Y}_{CMax}(p^U)) = (-\infty, +\infty)$ but
$[\hat{Y}_{AMin}(p^U), \hat{Y}_{AMax}(p^U)]$ is not, then it is as
if only the $G_{AP}$ part algorithm is in action;
\item[(c)] $(\hat{Y}_{AMin}(p^U), \hat{Y}_{AMax}(p^U)) = (-\infty, +\infty)$ but
$[\hat{Y}_{CMin}(p^U), \hat{Y}_{CMax}(p^U)]$ is not, then it is as
if only the $G_{P}$ part algorithm is in action;
\item[(d)] Neither window is $(-\infty, +\infty)$, we will have a
detailed discussion for the three algorithms: $G$-SSBa, $G$-ASIter
and $G$-AdvIter respectively.
\end{enumerate}

A general criterion is used to combine $G$-SSBa, $G$-ASIter and
$G$-AdvIter, this criterion is that we always give higher
priority to the window derived from $G_P$ than from $G_{AP}$. This
is because the former is more reliable than the latter, as the
latter may involve inaccuracy in information propagation.

\subsubsection{Algorithm for MYE in $G$: $G$-SSBa and $G$-ASIter}
Since the structures of $G$-SSBa and $G$-ASIter are similar, we try
to merge their pseudo codes together for space saving and ease of
description\footnote{In real implementation, they are separated.}.
The pseudo code of $G$-SSBa and $G$-ASIter for case (d) is listed in
Algorithm~\ref{Alg:GSSBaASIter}.
\begin{algorithm}
\caption{The pseudo code of $G$-SSBa and $G$-ASIter for case (d)}
\label{Alg:GSSBaASIter}
\begin{algorithmic}[1]
\IF{$G$-SSBa}
\STATE $[\hat{Y}_{CMin}, \hat{Y}_{CMax}] \leftarrow$ {\bf Simple
Window Derivation Method} in Eq.~(\ref{eq:y_cmin}) and
(\ref{eq:y_cmax});
\ELSIF{$G$-ASIter}
\STATE $[\hat{Y}_{CMin}, \hat{Y}_{CMax}] \leftarrow$ {\bf Advanced
Window Derivation Method} in Algorithm~\ref{Alg:GPAS};
\ENDIF
\REPEAT
    \STATE $[\hat{Y}_{AMin}, \hat{Y}_{AMax}] \leftarrow$ by
    $G_{AP}$-Ba, Eqs.~(\ref{eq:awmin}), (\ref{eq:awmax}), (\ref{eq:acwmin}),
    (\ref{eq:acwmax});
    \FORALL{$p^U \in V_P^U$}
    \STATE /* Init */
    \STATE $\hat{Y}_{GMin}(p^U) \leftarrow -\infty$;
    \STATE $\hat{Y}_{GMax}(p^U) \leftarrow +\infty$;
        \IF{$\hat{Y}_{CMin}(p^U) > -\infty$ and $\hat{Y}_{CMax}(p^U) < +\infty$}
            \STATE /* Type-1 Window in $G_P$ */
            \IF {$\hat{Y}_{AMin}(p^U) < \hat{Y}_{CMin}(p^U)$ or $\hat{Y}_{AMax}(p^U) > \hat{Y}_{CMax}(p^U)$}
                \STATE $\hat{Y}_{GMin}(p^U) \leftarrow \hat{Y}_{CMin}(p^U)$;
                \STATE $\hat{Y}_{GMax}(p^U) \leftarrow \hat{Y}_{CMax}(p^U)$;
            \ELSE
                \STATE $\hat{Y}_{GMin}(p^U) \leftarrow \max\{\hat{Y}_{CMin}(p^U), \hat{Y}_{AMin}(p^U)\}$;
                \STATE $\hat{Y}_{GMax}(p^U) \leftarrow \min\{\hat{Y}_{CMax}(p^U), \hat{Y}_{AMax}(p^U)\}$;
            \ENDIF
        \ELSIF{$\hat{Y}_{CMin}(p^U) > -\infty$ and $\hat{Y}_{CMax}(p^U) = +\infty$}
            \STATE /* Type-2 Window in $G_P$ */
            \IF {$\hat{Y}_{AMax}(p^U) < \hat{Y}_{CMin}(p^U)$}
                \STATE $\hat{Y}_{GMin}(p^U) \leftarrow \hat{Y}_{CMin}(p^U)$;
                \STATE $\hat{Y}_{GMax}(p^U) \leftarrow \hat{Y}_{CMin}(p^U)$;
            \ELSE
                \STATE $\hat{Y}_{GMin}(p^U) \leftarrow \max\{\hat{Y}_{CMin}(p^U), \hat{Y}_{AMin}(p^U)\}$;
                \STATE $\hat{Y}_{GMax}(p^U) \leftarrow \hat{Y}_{AMax}(p^U)$;
            \ENDIF
        \ELSIF{$\hat{Y}_{CMin}(p^U) = -\infty$ and $\hat{Y}_{CMax}(p^U) < +\infty$}
            \STATE /* Type-3 Window in $G_P$ */
            \IF {$\hat{Y}_{AMin}(p^U) > \hat{Y}_{CMax}(p^U)$}
                \STATE $\hat{Y}_{GMin}(p^U) \leftarrow \hat{Y}_{CMax}(p^U)$;
                \STATE $\hat{Y}_{GMax}(p^U) \leftarrow \hat{Y}_{CMax}(p^U)$;
            \ELSE
                \STATE $\hat{Y}_{GMin}(p^U) \leftarrow \hat{Y}_{AMin}(p^U)$;
                \STATE $\hat{Y}_{GMax}(p^U) \leftarrow \min\{\hat{Y}_{CMax}(p^U), \hat{Y}_{AMax}(p^U)\}$;
            \ENDIF
        \ELSE
            \STATE /* Type-4 Window in $G_P$ */
            \STATE Case (b);
        \ENDIF
        \STATE /* Simple Year Value Calculation */
        \STATE $\hat{Y}(p^U) \leftarrow \frac{\hat{Y}_{GMin}(p^U) +
        \hat{Y}_{GMax}(p^U)}{2}$;
    \ENDFOR
    \IF{$G$-SSBa}
    \STATE Break;
    \ENDIF
\UNTIL{No update happens}
\end{algorithmic}
\end{algorithm}

In Algorithm~\ref{Alg:GSSBaASIter}, we denote $\hat{Y}_{GMin}(p^U),
\hat{Y}_{GMax}(p^U)$ to be the two bounds of the derived year
estimation window in $G$. In the beginning, we derive
$[\hat{Y}_{CMin}, \hat{Y}_{CMax}]$ by a simple window derivation
method for algorithm $G$-SSBa, or an advanced window derivation method
for algorithm $G$-ASIter (Lines 1-5).

Next, we derive $[\hat{Y}_{GMin}, \hat{Y}_{GMax}]$ depending on the
type of the window in $G_P$, e.g., Lines 12-20 for Type-1, Lines
21-29 for Type-2 and Lines 30-38 for Type-3. The derivation follows
the general criterion that if the intersection of
$[\hat{Y}_{CMin}(p^U), \hat{Y}_{CMax}(p^U)]$ and
$[\hat{Y}_{AMin}(p^U), \hat{Y}_{AMax}(p^U)]$ is not empty, we take
this intersection window as $[\hat{Y}_{GMin}(p^U),
\hat{Y}_{GMax}(p^U)]$; otherwise, we take $[\hat{Y}_{CMin}(p^U),
\hat{Y}_{CMax}(p^U)]$. Line 44 is the same simple year value
calculation method as in $G_P$-SS, $G_P$-AS, $G_{AP}$-Ba and
$G_{AP}$-Iter. In fact, if conditions (Line 23 or Line 32) happen
(i.e. the two windows do not intersect with each other), the
operation (Lines 24-25 and Lines 33-34, together Line 44) is
equivalent to Eq.~(\ref{eq:CWestType2})-Eq.~(\ref{eq:CWestType3}),
taking the bound values. For $G$-SSBa of which the combination
includes $G_{AP}$-Ba, the basic procedure will only be carried out once
(Lines 46-48); While for $G$-ASIter of which the combination
includes $G_{AP}$-Iter, the $[\hat{Y}_{GMin}, \hat{Y}_{GMax}]$
window will be propagated until convergence (Line 6 together with
Line 49).

\subsubsection{Algorithm for MYE in $G$: $G$-AdvIter}
$G$-AdvIter is the combination of $G_P$-AA and $G_{AP}$-AdvIter,
therefore, the concepts of training set $\mathcal{T}$ as well as the
Consistent-Coauthor-Pair relationship will be involved.
Algorithm~\ref{Alg:GAdvIter} list the pseudo code of $G$-AdvIter for
case (d):
\begin{algorithm}
\caption{The pseudo code of $G$-AdvIter for case (d)}
\label{Alg:GAdvIter}
\begin{algorithmic}[1]
\STATE Run Algorithm~\ref{Alg:GPAA}, derive $[\hat{Y}_{CMin},
\hat{Y}_{CMax}]$ and the training set $\mathcal{T}$.
 \FORALL{$p^U \in V_P^U$}
    \STATE Derive the Consistent-Coauthor-Pair set, $\Omega(p^U)$.
 \ENDFOR
\REPEAT
    \STATE $[\hat{Y}_{AMin}, \hat{Y}_{AMax}] \leftarrow$ by
    $G_{AP}$-Ba, Eqs.~(\ref{eq:awmin}), (\ref{eq:awmax}), (\ref{eq:acwmin}),
    (\ref{eq:acwmax});
    \FORALL{$p^U \in V_P^U$}
    \STATE$ \hat{Y}(p^U) \leftarrow Null$;
        \IF{$\hat{Y}_{CMin}(p^U) > -\infty$ and $\hat{Y}_{CMax}(p^U) < +\infty$}
            \STATE /* Type-1 Window in $G_P$ */
            \STATE Derivation of $\hat{Y}_{GMin}(p^U),
            \hat{Y}_{GMax}(p^U)$; /* Same as
            Algorithm~\ref{Alg:GSSBaASIter}, Lines 12-20 */
            \STATE /* Year Value Calculate */
            \IF{$\Omega(p^U) \cap V_P^K \neq \emptyset$}
                \STATE $\hat{Y}(p^U) \leftarrow W_G(p^U, \gamma, \hat{Y}_{CMin}(p^U), \hat{Y}_{CMax}(p^U))$
            \ENDIF
            \IF[In case $W_G$ does not work]{$\hat{Y}(p^U) = Null$}
                \STATE $\hat{Y}(p^U) \leftarrow \frac{\hat{Y}_{GMin}(p^U) +
        \hat{Y}_{GMax}(p^U)}{2}$;
            \ENDIF
        \ELSIF{$\hat{Y}_{CMin}(p^U) > -\infty$ and $\hat{Y}_{CMax}(p^U) = +\infty$}
            \STATE /* Type-2 Window in $G_P$ */
            \STATE Derivation of $\hat{Y}_{GMin}(p^U),
            \hat{Y}_{GMax}(p^U)$; /* Same as
            Algorithm~\ref{Alg:GSSBaASIter}, Lines 21-29 */
            \STATE /* Year Value Calculate */
            \STATE $dResult \leftarrow d(\textrm{Type-2},
            \hat{Y}_{CMin}(p^U))$;/*call $d(WinType(p^U),
            BoundVal(p^U)$*/
            \STATE $\delta \leftarrow dResult-\hat{Y}_{CMin}(p^U)$;
            \IF{$\Omega(p^U) \cap V_P^K \neq \emptyset$}
                \STATE $\hat{Y}(p^U) \leftarrow W_G(p^U, \gamma, \hat{Y}_{CMin}(p^U), \hat{Y}_{CMin}(p^U)+2\delta)$
            \ENDIF
            \IF[In case $W_G$ does not work]{$\hat{Y}(p^U) = Null$}
                \IF{$\hat{Y}_{AMax}(p^U) < \hat{Y}_{CMin}(p^U)$ or $dResult \in (\hat{Y}_{GMin}(p^U), \hat{Y}_{GMax}(p^U))$}
                    \STATE $\hat{Y}(p^U) \leftarrow dResult$
                \ELSE
                    \STATE $\hat{Y}(p^U) \leftarrow \frac{\hat{Y}_{GMin}(p^U) + \hat{Y}_{GMax}(p^U)}{2}$;
                \ENDIF
            \ENDIF
        \ELSIF{$\hat{Y}_{CMin}(p^U) = -\infty$ and $\hat{Y}_{CMax}(p^U) < +\infty$}
            \STATE /* Type-3 Window in $G_P$ */
             \STATE Derivation of $\hat{Y}_{GMin}(p^U),
            \hat{Y}_{GMax}(p^U)$; /* Same as
            Algorithm~\ref{Alg:GSSBaASIter}, Lines 30-38 */
            \STATE /* Year Value Calculate */
            \STATE $dResult \leftarrow d(\textrm{Type-3},
            \hat{Y}_{CMax}(p^U))$;/*call $d(WinType(p^U),
            BoundValue(p^U)$*/
            \STATE $\delta \leftarrow \hat{Y}_{CMax}(p^U)-dResult$;
            \IF{$\Omega(p^U) \cap V_P^K \neq \emptyset$}
                \STATE $\hat{Y}(p^U) \leftarrow W_G(p^U, \gamma, \hat{Y}_{CMax}(p^U)-2\delta, \hat{Y}_{CMax}(p^U))$
            \ENDIF
            \IF[In case $W_G$ does not work]{$\hat{Y}(p^U) = Null$}
                \IF{$\hat{Y}_{AMin}(p^U) > \hat{Y}_{CMax}(p^U)$ or $dResult \in (\hat{Y}_{GMin}(p^U), \hat{Y}_{GMax}(p^U))$}
                    \STATE $\hat{Y}(p^U) \leftarrow dResult$
                \ELSE
                    \STATE $\hat{Y}(p^U) \leftarrow \frac{\hat{Y}_{GMin}(p^U) + \hat{Y}_{GMax}(p^U)}{2}$;
                \ENDIF
            \ENDIF
        \ELSE
            \STATE /* Type-4 Window in $G_P$: Case (b); */
        \ENDIF
    \ENDFOR
\UNTIL{No update happens}
\end{algorithmic}
\end{algorithm}

In Algorithm~\ref{Alg:GAdvIter}, we omit the same code of deriving
$\hat{Y}_{GMin}(p^U), \hat{Y}_{GMax}(p^U)$ as in
Algorithm~\ref{Alg:GSSBaASIter} (Lines 11, 21, 37). At the beginning
(Line 1), we call the function $G_P$-AA (Algorithm~\ref{Alg:GPAA})
to derive $\hat{Y}_{CMin}(p^U), \hat{Y}_{CMax}(p^U)$ and the
training set $\mathcal{T}$, which is a series of 3-tuple data
$\{y_{t}, WinType_{t}, BoundVal_{t}\}$ from the papers with known
year information. The preparation of the Consistent-Coauthor-Pair
set for each missing year paper $\Omega(p^U)$, like
$G_{AP}$-AdvIter, is also called (Lines 2-4). The main difference
between $G$-AdvIter and $G$-ASIter is the method of calculating year
value. For all three types of window in $G_P$, we apply the
$W_G(p^U, \gamma, y_l, y_r)$ function to calculate the year value:
\begin{eqnarray}
\textrm{When}\;\Omega_G(p^U) &=& \{q|q\in\Omega(p^U), Y(q)\in(y_l,
y_r)\},\textrm{and}\;\Omega_G(p^U)\cap V_P^K
\neq\emptyset,\nonumber\\
W_G(p^U,\gamma, y_l, y_r) &=& \frac{\sum_{q\in \Omega_G(p^U)\cap
V_P^K}\;\;w(p^U, q)^\gamma \times Y(q)}{\sum_{q\in \Omega_G(p^U)\cap
V_P^K} w(p^U, q)^\gamma};\nonumber\\
\textrm{Otherwise}&=& Null. \label{eq:wG}
\end{eqnarray}

In Eq.~(\ref{eq:wG}), the different part of $W_G$ is that we pick
out a subset of papers from $\Omega(p^U)$, denoted by
$\Omega_G(p^U)$, satisfying the condition that the paper publishing
years are within an input window $[y_l, y_r]$, i.e., $\Omega_G(p^U)
=\{q|q\in\Omega(p^U), Y(q)\in[y_l, y_r]\}$. For Type-1 window of
$G_P$, we choose the subset $\Omega_G(p^U)$ by setting the input
window to be $[y_l=\hat{Y}_{CMin}(p^U), y_r=\hat{Y}_{CMax}(p^U)]$
for calculating $\hat{Y}(p^U)$ (Line 14). But if $\Omega_G(p^U)\cap
V_P^K = \emptyset$, we revert back to the default way (Lines 16-18).

The process for a Type-2 or Type-3 window is a little more
complicated. For a Type-2 window, both $\Omega(p^U)$ and $\mathcal{T}$
are available tools. Due to this we propose the following way: we first derive
the estimation year value, denoted by $dResult$, through $d(\cdot)$
function expressed in Eq.~(\ref{eq:d_func}). We use this $dResult$
and the input parameter $\hat{Y}_{CMin}(p^U)$ to define a window
[$y_l=\hat{Y}_{CMin}(p^U)$, $y_r=\hat{Y}_{CMin}(p^U) + 2\delta]$, of
which the interval equals to twice of the distance from $dResult$ to
$\hat{Y}_{CMin}(p^U)$, $\delta = dResult - \hat{Y}_{CMin}(p^U)$.
This window is then used to derive $\Omega_G(p^U)$ and calculate
$\hat{Y}(p^U)$ (Lines 23-27). If $\Omega_G(p^U)\cap V_P^K =
\emptyset$, we have a second choice which is $dResult$, if one of
the following two conditions is met: (a) The two windows
$[\hat{Y}_{CMin}(p^U), \hat{Y}_{CMax}(p^U)]$ and
$[\hat{Y}_{AMin}(p^U), \hat{Y}_{AMax}(p^U)]$ have no intersection;
or (b) $dResult \in [\hat{Y}_{GMin}(p^U), \hat{Y}_{GMax}(p^U)]$
(Lines 28-31). Otherwise, we change back to the default way (Line
32).

The process for a Type-3 window is symmetric to Type-2. The only
difference is that the input window for deriving $\Omega_G(p^U)$ and
$\hat{Y}_{CMin}(p^U)$ becomes $W_G(p^U, \gamma,
y_l=\hat{Y}_{CMax}(p^U)-2\delta, y_r=\hat{Y}_{CMax}(p^U))$ (Lines
39-43).


%
%
%
\section{Experiment Results}\label{Sec:exp}
In this section we present the experiment settings and evaluation
results. In the experiment, we test the proposed MYE algorithms in
the last section by applying them to all the three types of the
academic social networks, the paper citation network $G_P$, the
paper authorship network $G_{AP}$ and the heterogenous network $G$.
\subsection{Data Sets}
We have tried three different data sets: Microsoft academic data
set~\cite{libra}, DBLP~\cite{dblp} with additional citation
information, DBLP-Cit data set~\cite{Tang:07ICDM,Tang:08KDD}, and
American Physical Society data set~\cite{apsdataset}. The raw data
sets are not perfect in that: (a) there exits a proportion of
missing year papers; (b) Some citation links are pointing from early
published papers to later ones, which
breaks Assumption~\ref{assumption1}.

Since the performance evaluation needs ground truth knowledge, we
have to do some preprocessing on the original data sets, including:
a) remove these missing year papers and their relationships
(citation links and paper-authorship links); b) remove those
citation links that break Assumption~\ref{assumption1}.

Table \ref{Tab:datasets} lists the general information about the
three data sets after preprocessing:
\begin{table}[htb]
\centering
\begin{tabular}{|c|c|c|c|}
\hline    Data set         & Microsoft Libra     & DBLP-Cit      & APS \\
\hline    Input Window     & (1900 - 2013)       & (1900 - 2013) & (1900 - 2013)\\
\hline    \#papers         & 2323235             & 1558503       & 463347\\
\hline    \#authors        & 1278407             & 914053        & 320964\\
\hline    \#total citation links & 10003121     & 2062896       & 4689142\\
\hline
\end{tabular}
\caption{General information of the three data sets used after
preprocessing.} \label{Tab:datasets}
\end{table}

As we can see in Table~\ref{Tab:datasets}, the average number of
citation links per paper of the three data sets are: 4.31 for Libra,
1.33 for DBLP-Cit and 10.34 for APS, which appears disparate. This
probably reflects how well these three data sets are collected and
managed. The APS data set is the most complete in terms of the paper
citation information, and the DBLP-Cit is probably the
least\footnote{DBLP~\cite{dblp} is a popular and well-managed data
set, with complete and accurate meta information. But it does not
provide paper citation information. DBLP-Cit is created based on the
original DBLP paper set with adding paper citation relationships
through proper mining method~\cite{Tang:07ICDM,Tang:08KDD}}. For
DBLP-Cit, the job to find citation links for an existing paper set
is a big challenge. The small number of average paper citation links
shows that it is likely only a small proportion of the complete paper
citation links are found.

The completeness and accuracy of the citation links will only affect
those MYE algorithms that rely on citation information, e.g., the
three algorithms for $G_P$.

\subsection{Evaluation methodology}
We apply a similar approach to the K-fold cross validation
method~\cite{mosteller1968data,kohavi1995study} to evaluate the MYE
algorithms. For each date set after pre-processing, we randomly
split the paper set into $K$ mutually exclusive groups, i.e., $V_P =
\bigcup_{k=1}^K V_{P_k}, \textrm{and}\;\forall i\neq j, V_{P_i}\cap
V_{P_j} = \emptyset$. In addition, each group has approximately the
same size, $|V_{P_k}| \approx \frac{|V_P|}{K}, k = 1, 2, \dots, K$.

For a given parameter $K$, the experiment repeats $K$ times. In the
$j$th time, the year information of the papers in group $V_{P_j}$ is
artificially hidden, thus assumed to be the missing year paper set
$V_P^U = V_{P_j}$, and the remaining groups become the paper set
with known year information, i.e., $V_P^K = V_P\setminus V_{P_j}$.
The overall performance metrics take the average of the results
obtained in each of the $K$ times.

Indirectly, the value of $K$ controls the severity of the missing
year phenomenon. For convenience, we define $\eta =
\frac{|V_P^U|}{|V_P|} \approx \frac{1}{K}$ to be the \emph{Missing
Year Ratio} of the data set. Throughout the experiment, we have
tried 5 different $\eta = \frac{1}{8}, \frac{1}{5}, \frac{1}{4},
\frac{1}{3}, \frac{1}{2}$.

\subsection{Performance metrics}
Three metrics are used to evaluate the performance of the MYE
algorithms.
\begin{enumerate}
\item[1)] Coverage

We have defined the uncovered ratio in Section~\ref{Sec:method}. It
equals to the number of those missing year papers finally labeled as
\emph{Uncovered} by MYE algorithms, divided by the total number of
missing year papers $|V_P^U|$. We use $N^U = |V_P^U| -
\textrm{Total\#Uncovered}$ to denote the number of the covered part.
In one experiment, the coverage metric is equal to
$\frac{N^U}{|V_P^U|}$. With K-fold cross validation, the overall
coverage becomes:
\begin{equation}
Coverage = \frac{1}{K}\sum_{k=1}^K \frac{N^U_k}{|V_{P_k}|},
\end{equation}
where the subscript $k$ indicates the $k$th iteration and $V_P^U =
V_{P_k}$.

\item[2)] Mean absolute error (MAE)
\begin{equation}
MAE = \frac{1}{K}\sum_{k=1}^K \bigg(
\frac{1}{N^U_k}\sum_{i=1}^{N^U_k} |Y(p^U_i) - \hat{Y}(p^U_i)|\bigg),
\end{equation}
where the $k$th iteration, $V_P^U = V_{P_k}$, $\hat{Y}(p^U_i)$ is
the estimated year. $Y(p^U_i)$ is the real year of $p^U_i$, which we
assumed to be unknown when running the MYE algorithms and used only
for validation purposes.

\item[3)] Root mean square error (RMSE)
\begin{equation}
RMSE = \frac{1}{K}\sum_{k=1}^K
\bigg(\sqrt{\frac{1}{N^U_k}\sum_{i=1}^{N^U_k} \big[Y(p^U_i) -
\hat{Y}(p^U_i)\big]^2}\;\;\bigg).
\end{equation}
\end{enumerate}

In order to have a better understanding of the coverage metric, we
propose an analytical model to calculate the expected coverage for
an undirected graph $G = (V, E)$. According to the basic graph
theory~\cite{kleinburgNetwork}, $G$ can be partitioned into $S$
connected components $G = \bigcup_{i=1}^S G_i$, where $\forall i, j,
G_i \cap G_j = \emptyset$.

The iteration mechanism of the MYE algorithms (e.g., $G_{AP}$-Iter,
or $G_{AP}$-AdvIter) ensures that there can be only two possible
outcomes for any connected component $G_i = (V_i, E_i)$ when
propagation stops\footnote{The outcome of $G_P$-AS and $G_P$-AA is a
little complicated, we will discuss it later.}:

(I) All the missing year papers in this component have feasible
estimated values (hence, $\neq$ \emph{Uncovered}), if and only if
there exits at least one paper with known year information in this
component, i.e., $V_i \cap V_P^K \neq \emptyset$;

(II) Otherwise, all the missing year papers in this component are
labeled as \emph{Uncovered}.

If we assume the missing year paper is uniformly distributed among
the whole paper set, then the expected coverage value can be
calculated by Eq.~(\ref{Eq:ExpCoverage}):
\begin{equation}\label{Eq:ExpCoverage}
 Coverage\big(\eta, \bigcup_{i=1}^S V_i\big) = 1 -
\frac{\sum_{i=1}^S\; \eta^{|V_i|}\cdot|V_i|}{\eta|V|}\;.
\end{equation}

In Eq.~(\ref{Eq:ExpCoverage}), there are two inputs for this
calculation: the year missing ratio $\eta$ and the vertex partition
$V = \bigcup_{i=1}^S V_i$. According to the uniform distribution
assumption, each paper is selected to be the missing year paper with
equal probability $\eta$. Thus the denominator equals to the
expected number of missing year papers $|V_P^U| = \eta|V|$. For each
component $G_i$, $\eta^{|V_i|}$ is the probability that all the
papers in it are missing year papers and $\eta^{|V_i|}\cdot|V_i|$ is
hence the expected number of papers that will be labeled as
\emph{Uncovered}.

For the three types of the academic social networks, the above model
actually cannot be applied directly. To apply it, we have to make
proper modifications: (1) based on the citation network $G_P = (V_P,
E_P)$, we construct $G_P'= (V_P, E_P')$ by implicitly considering
all the citation edges as undirected edges, where $E_P'$ is the
undirected edge set. (2) based on the paper authorship network
$G_{AP} = (V_A \cup V_P, E_{AP})$, we build a coauthor indicator
graph $G_{AP}' = (V_P, E_{PP})$, where the existence of an edge
between two papers in $G_{AP}'$ indicates that they have at least
one common author, i.e., $\forall e_{i,j} \in E_{PP}, i, j \in V_P
\Leftrightarrow A(i)\cap A(j) \neq \emptyset$, where $A(i)$ is the
author set of paper $i$. (3) For the heterogenous network $G$, by
simply combining $G_P'$ and $G_{AP}'$, we obtain $G' = (V_P,
E_P'\cup E_{PP})$. Now the analytical model can be applied on
$G_P'$, $G_{AP}'$ and $G'$ to calculate the expected coverage.

\subsection{Experiment results in the citation network $G_P$}
\begin{figure*}[hbt]
    \centering
    \subfigure[Coverage-Libra]{
    \label{Fig:GPCovLibra}
    \includegraphics[width=1.4in]{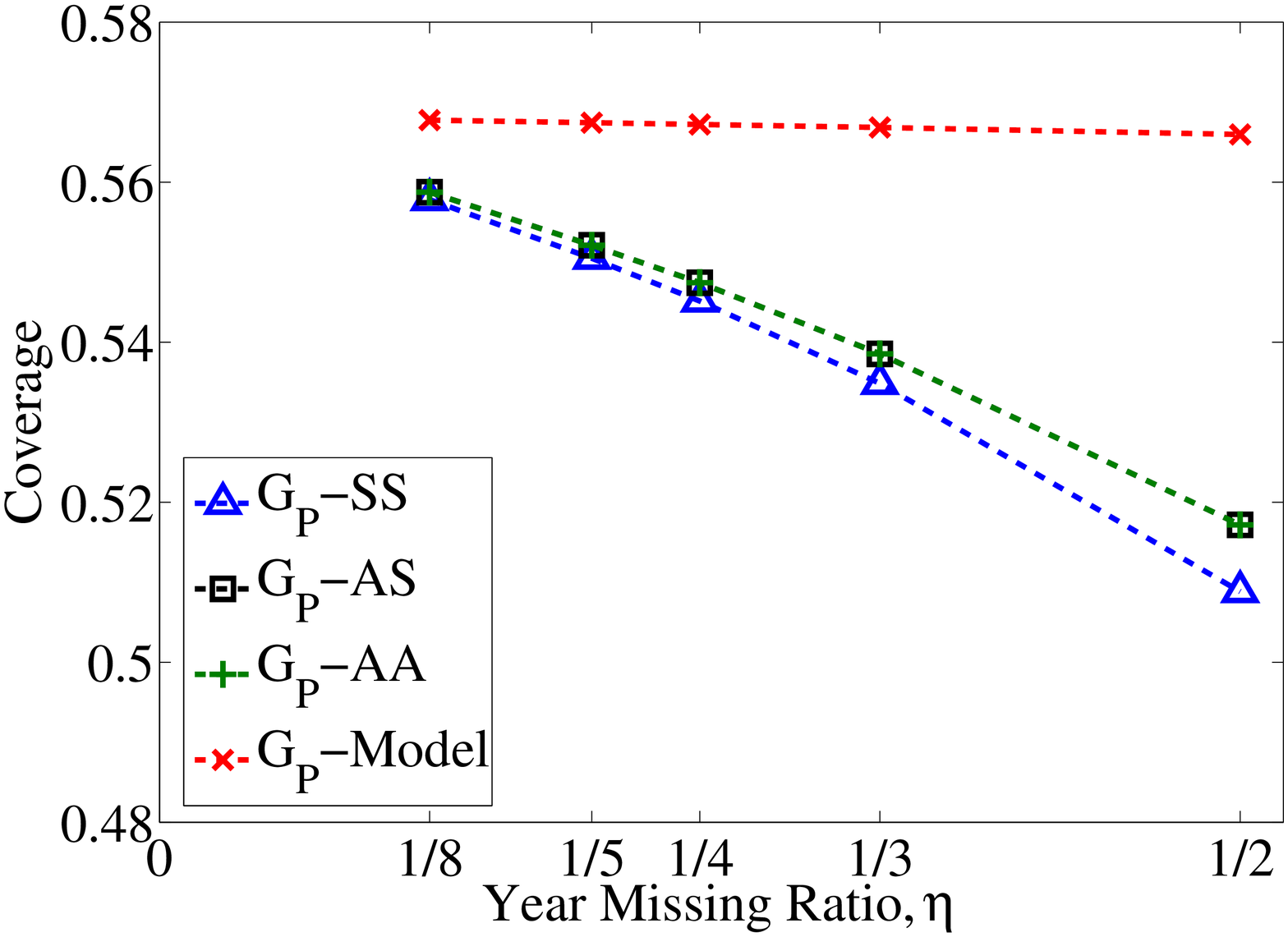}}
    \subfigure[Coverage-DBLP]{
    \label{Fig:GPCovDBLP}
    \includegraphics[width=1.4in]{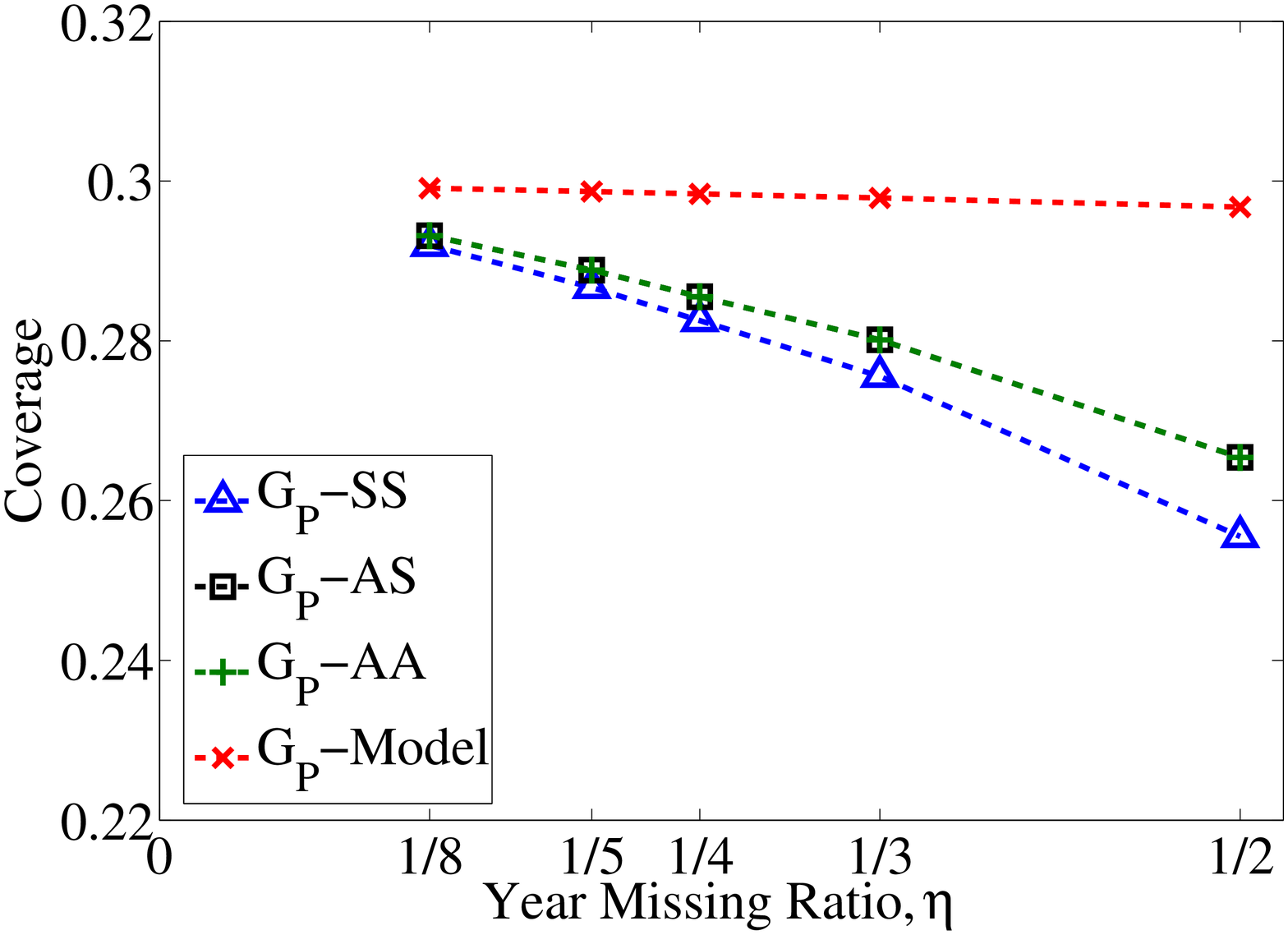}}
    \subfigure[Coverage-APS]{
    \label{Fig:GPCovAPS}
    \includegraphics[width=1.4in]{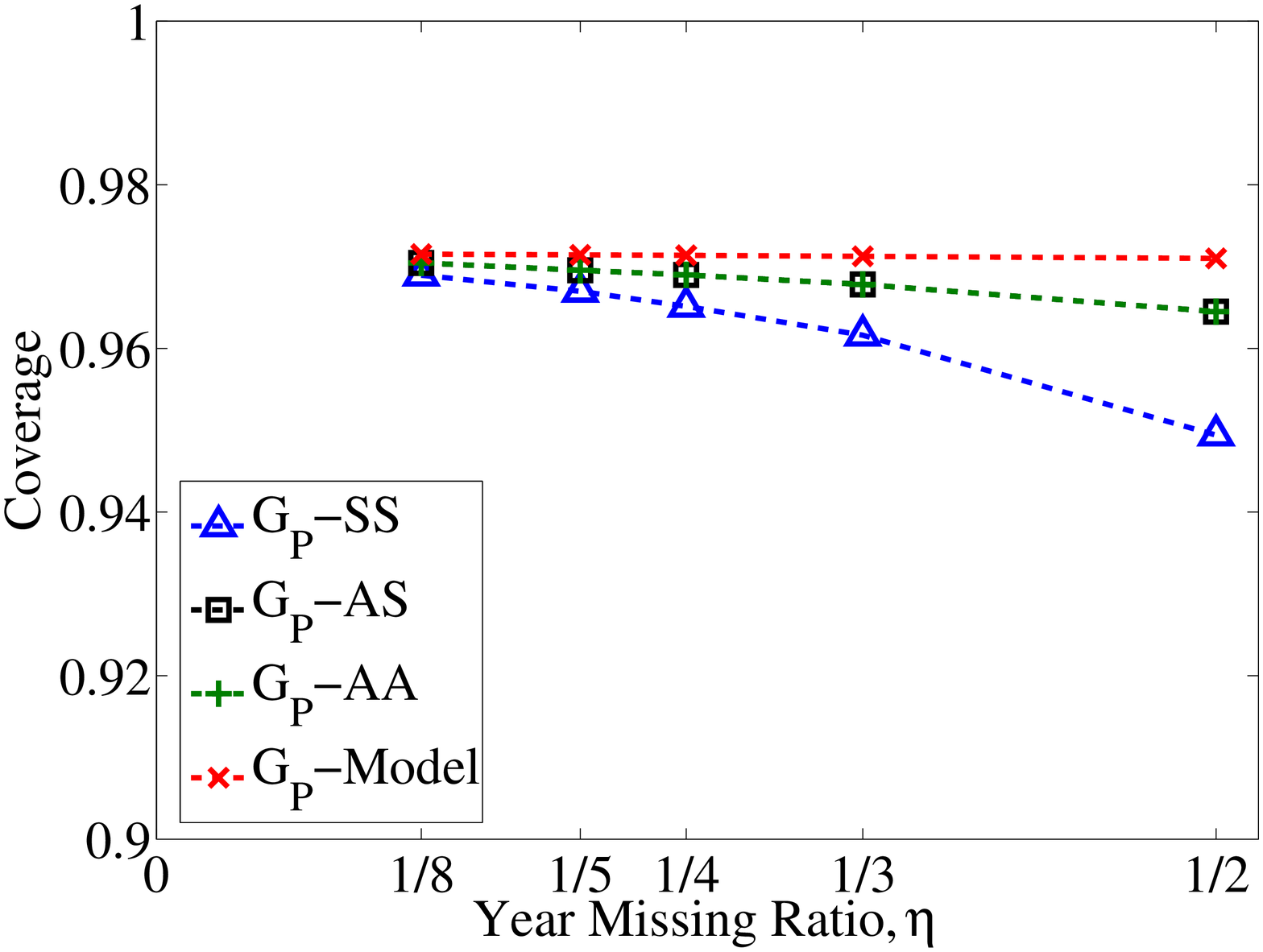}}
    \subfigure[MAE-Libra]{
    \label{Fig:GPMaeLibra}
    \includegraphics[width=1.4in]{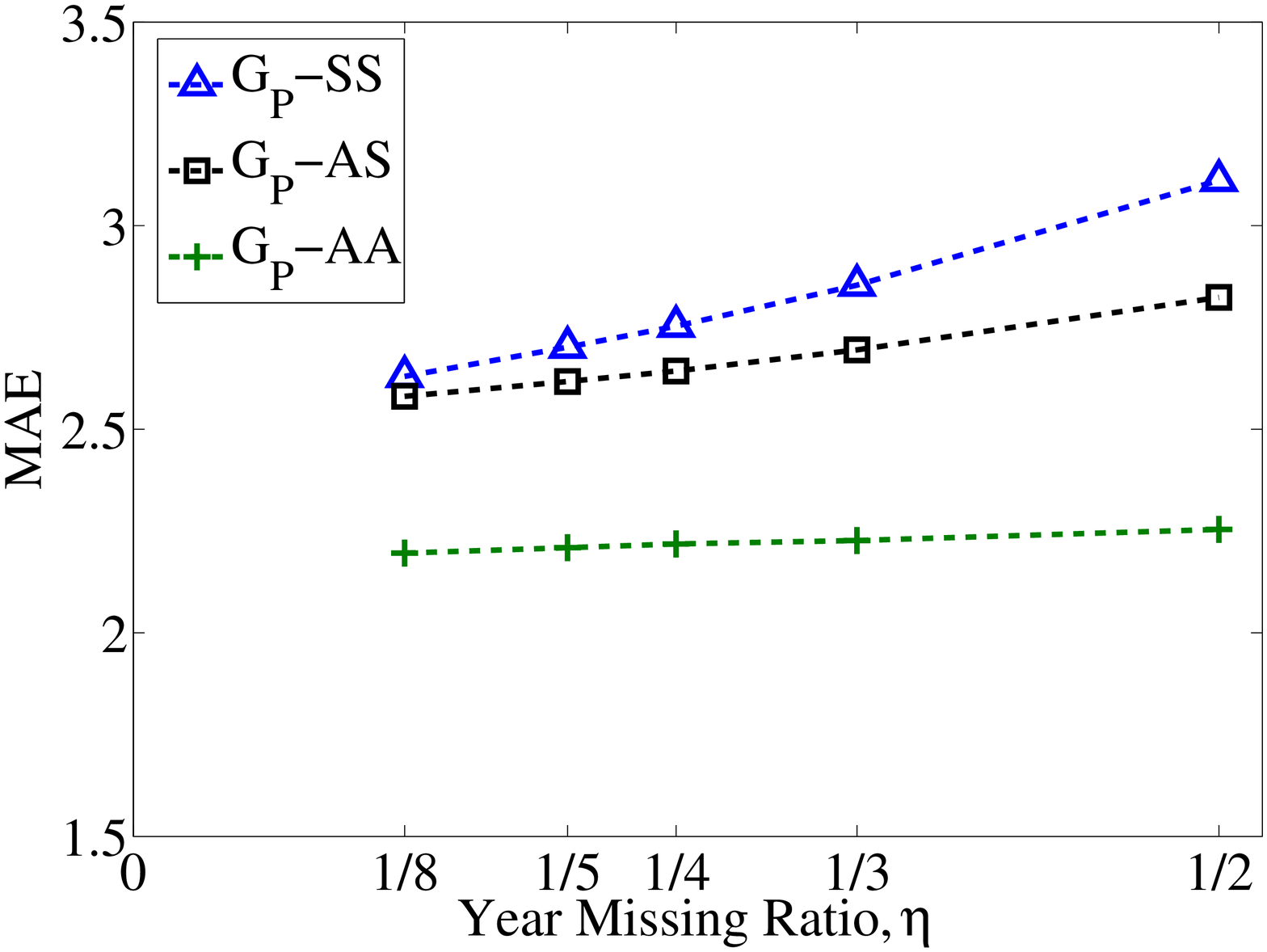}}
    \subfigure[MAE-DBLP]{
    \label{Fig:GPMaeDBLP}
    \includegraphics[width=1.4in]{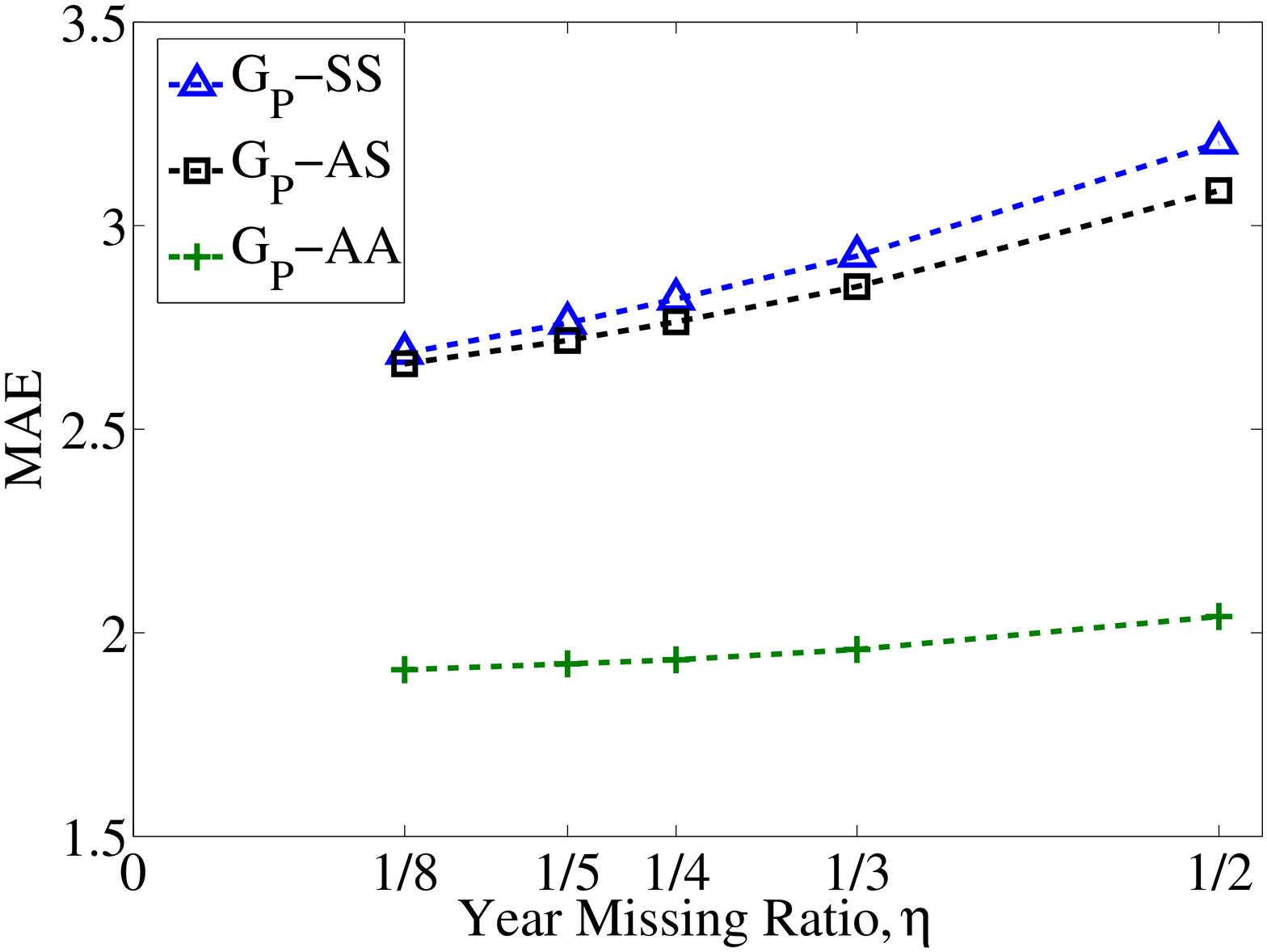}}
    \subfigure[MAE-APS]{
    \label{Fig:GPMaeAPS}
    \includegraphics[width=1.4in]{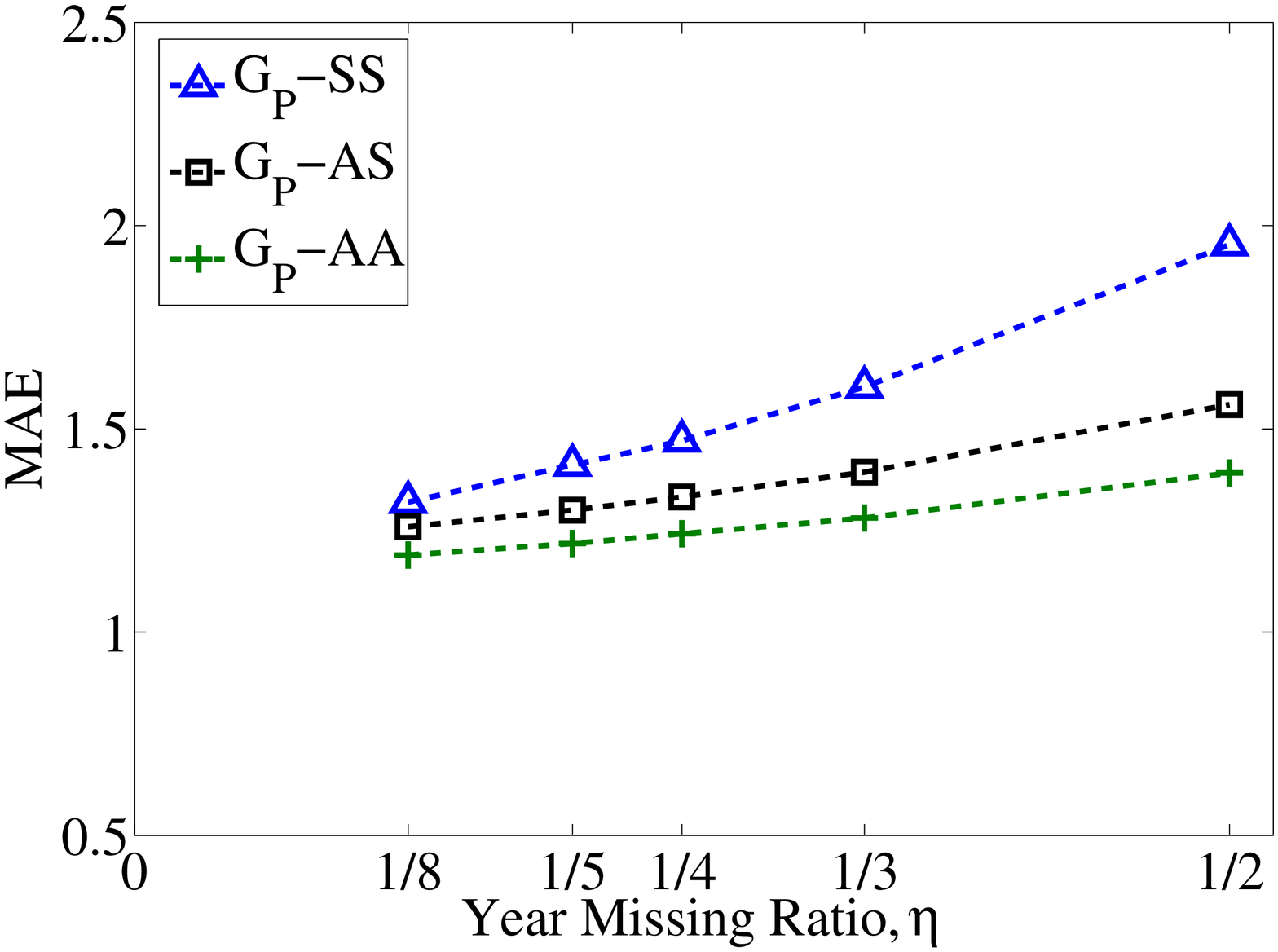}}
    \subfigure[RMSE-Libra]{
    \label{Fig:GPRmseLibra}
    \includegraphics[width=1.4in]{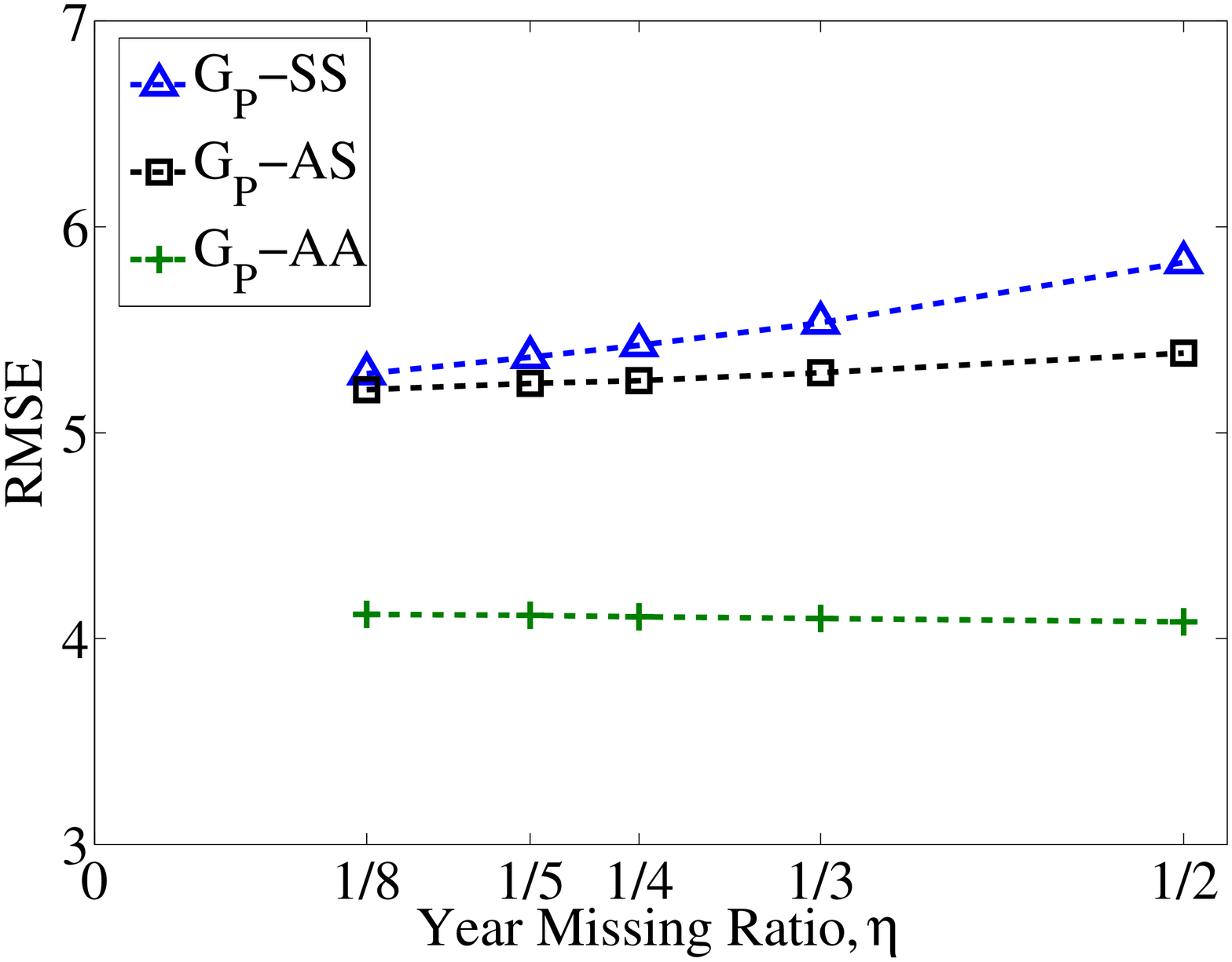}}
    \subfigure[RMSE-DBLP]{
        \label{Fig:GPRmseDBLP}
    \includegraphics[width=1.4in]{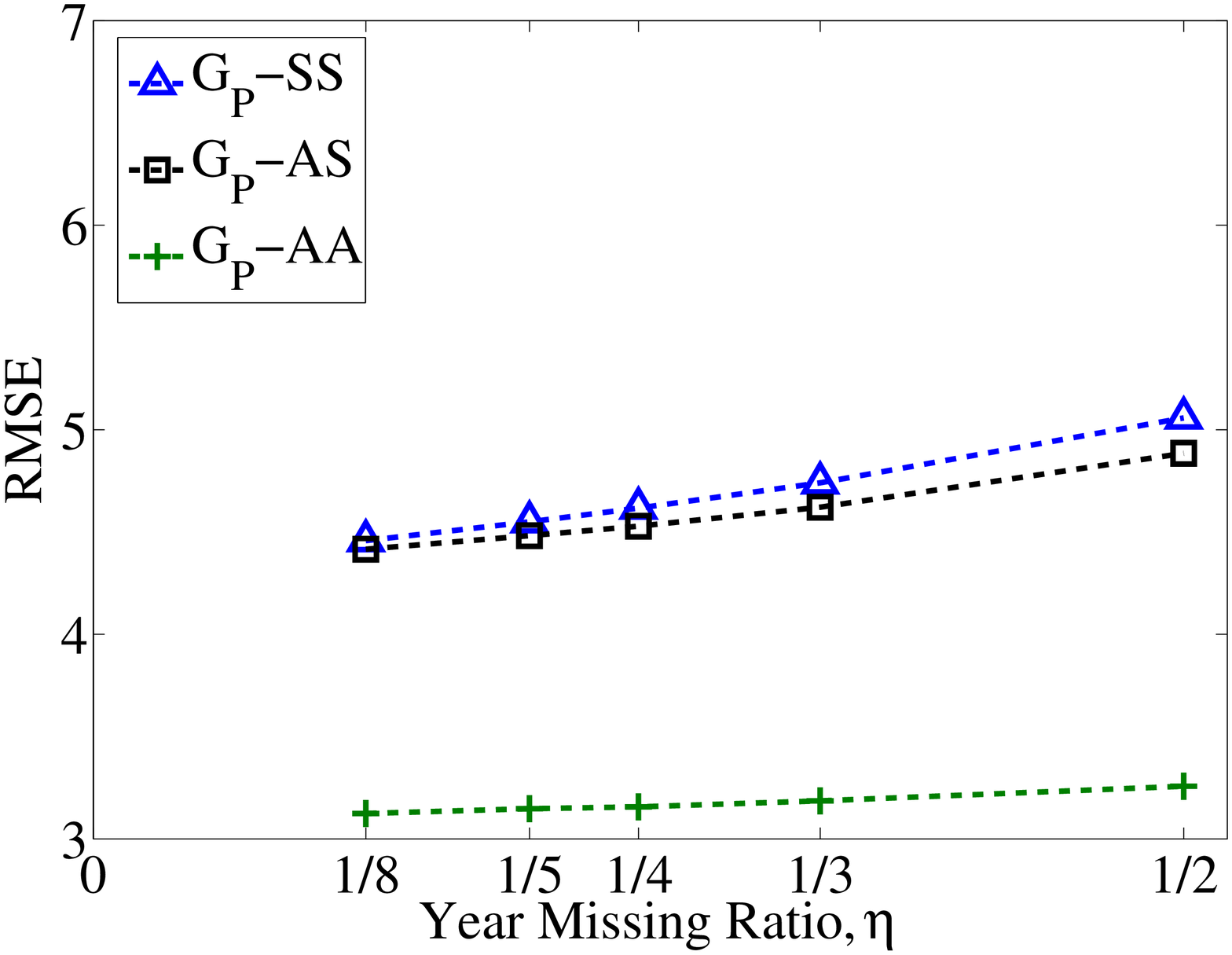}}
    \subfigure[RMSE-APS]{
    \label{Fig:GPRmseAPS}
    \includegraphics[width=1.4in]{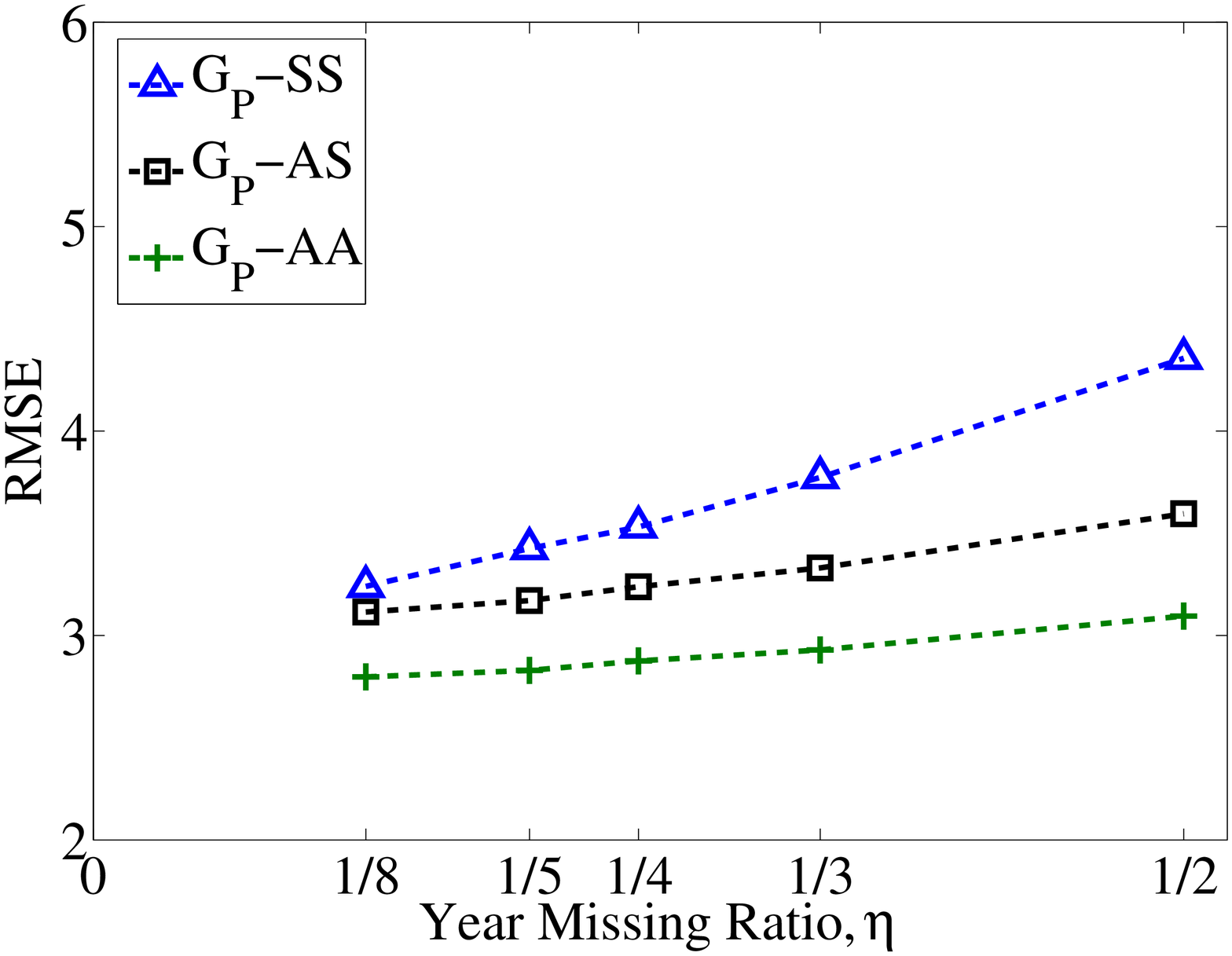}}
    %
    \caption{The Coverage, MAE and RMSE of algorithms $G_P$-SS (Simple Window Derivation and Simple Value Calculation),
    $G_P$-AS (Advanced Window Derivation and Simple Value Calculation) and
    $G_P$-AA (Advanced Window Derivation and Advanced Value Calculation) in a paper citation network $G_P$ of three data sets}
    \label{Fig:GPLibraDBLPAPS}
\end{figure*}

The first set of experiments are conducted on the citation network
$G_P = (V_P, E_P)$. The coverage, MAE and RMSE results of algorithms
$G_P$-SS, $G_P$-AS and $G_P$-AA are plotted in
Figure~\ref{Fig:GPLibraDBLPAPS}.

As shown in Figure~\ref{Fig:GPLibraDBLPAPS}, we have the following
observations:
\begin{enumerate}
\item[1)] For all the three algorithms, when $\eta$
increases, coverage decreases while both MAE and RMSE increase. This
implies the more information available helps to get better
estimation results, more coverage and less estimation error.

\item[2)] In Fig.~\ref{Fig:GPCovLibra}-\ref{Fig:GPCovAPS},
the curve of $G_P$-AS overlaps with that of $G_P$-AA and they have
better coverage than $G_P$-SS. This is consistent with what we have
discussed in Section~\ref{Sec:method} ($G_P$-AS and $G_P$-AA use the
same advanced window derivation method). However, it appears that
all the three coverage curves have certain deviation from the curve
(with nodes of red ``X'' in
Fig.~\ref{Fig:GPCovLibra}-\ref{Fig:GPCovAPS}) obtained by the
analytical model in
Eq.~(\ref{Eq:ExpCoverage}).\\
\\
The reason for this is that the analytical model overestimates the number of
covered papers for $G_P$-AS and $G_P$-AA. Recall that in
Section~\ref{Sec:method}, the window propagation method in $G_P$ is
different to the iteration scheme of $G_{AP}-Iter$ and
$G_{AP}-AdvIter$ in that it follows the bound transmission rules in
Eq.~(\ref{assumption1New}) and does not utilize estimation results
in the previous rounds. As a result, the outcome (I) discussed above
may not always be true, while (II) remains true. We use a typical
and simple example to illustrate. As shown in
Fig.~\ref{Eq:ExpException}, there are three papers ($a$, $b$, $c$)
and two citation links, where only one paper $b$ has year
information while the other two are missing year papers.
Fig.~\ref{Eq:ExpException} plots all the 7 possible topologies.
\begin{figure}[hbt]
    \centering
    \includegraphics[width=2.4in]{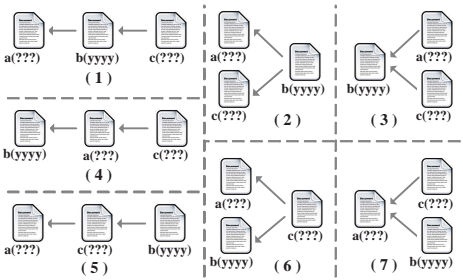}
    \caption{An example of paper citation network with three papers
    ($a, b, c$) and two citation links. When two papers are missing year
    ($a$ and $c$), there are totally 7 possible topologies.}
    \label{Eq:ExpException}
\end{figure}

According to outcome (I) of the analytical model, neither $a$ nor
$c$ will be labeled as \emph{Uncovered}. However, in
Fig.~\ref{Eq:ExpException} paper $a$ in case (6), and paper $c$ in
case (7) get an \emph{Uncovered} result by applying the advanced window
derivation method in Eq.~(\ref{assumption1New}). This would build a more
precise analytical model for the citation network, however this is too
complicated. Therefore, we stick to the current method in
Eq.~(\ref{Eq:ExpCoverage}) as an upper bound for the coverage
achieved in citation network $G_P$.

\item[3)] $G_P$-AA outperforms the other two for all network types
and data sets in terms of both coverage and estimation accuracy, MAE
and RMSE.

\item[4)] Comparing the three data sets, we find that the coverage
on APS data is much higher than the other two and DBLP-Cit is the
lowest. This is mainly caused by the completeness of the citation
information of the three data sets, mentioned in the beginning of
this section. Since APS maintains very complete and accurate
citation information, this benefits both coverage and accuracy for
the MYE in paper citation network (Fig.~\ref{Fig:GPCovAPS},
Fig.~\ref{Fig:GPMaeAPS}, Fig.~\ref{Fig:GPRmseAPS}).

\item[5)] In Fig.~\ref{Fig:GPCovLibra} and Fig.~\ref{Fig:GPCovDBLP},
the coverage on the Libra case is higher than DBLP-Cit, however, its MAE
and RMSE are at similar levels (or worse, e.g., $G_P$-AA in
Fig.~\ref{Fig:GPMaeLibra} and Fig.~\ref{Fig:GPMaeDBLP}, all the
curves in Fig.~\ref{Fig:GPRmseLibra} versus
Fig.~\ref{Fig:GPRmseDBLP}). We think one possible reason is that
quantitatively, Libra has a more complete paper citation information
than DBLP-Cit, but qualitatively, the correctness of Libra data may
be worse. We summarise this in Table~\ref{Tab:DataQuaGP}.
\begin{table}[htb]
\centering
\begin{tabular}{|c|c|}
\hline
\multicolumn{2}{|c|}{MYE performance in $G_P$}\\
\hline    Coverage    & APS $>$ Libra $>$ DBLP\\
\hline    MAE/RMSE         & APS $<$ DBLP $<$ Libra\\
\hline
\multicolumn{2}{|c|}{Inferred data quality of paper citation information}\\
\hline    Completeness & APS $>$ Libra $>$ DBLP\\
\hline    Correctness  & APS $>$ DBLP $>$ Libra\\
\hline
\end{tabular}
\caption{Summary on data quality of paper citation information of
three used datasets inferred from MYE performance in $G_P$.}
\label{Tab:DataQuaGP}
\end{table}
\end{enumerate}

\subsection{Experiment results for the paper authorship network $G_{AP}$}
The second set of experiments are conducted on the paper author
bipartite network $G_{AP} = (V_A \cup V_P, E_{AP})$. The coverage,
MAE and RMSE results of algorithms $G_{AP}$-Ba (the basic scheme),
$G_{AP}$-Iter (Simple iteration of the basic scheme) and
$G_{AP}$-AdvIter (Iteration with considering
Consistent-Coauthor-Pair information) are plotted in
Figure~\ref{Fig:GAPLibraDBLPAPS}. Our observations are:
\begin{figure*}[hbt]
    \centering
    \subfigure[Coverage-Libra]{
    \label{Fig:GAPCovLibra}
    \includegraphics[width=1.4in]{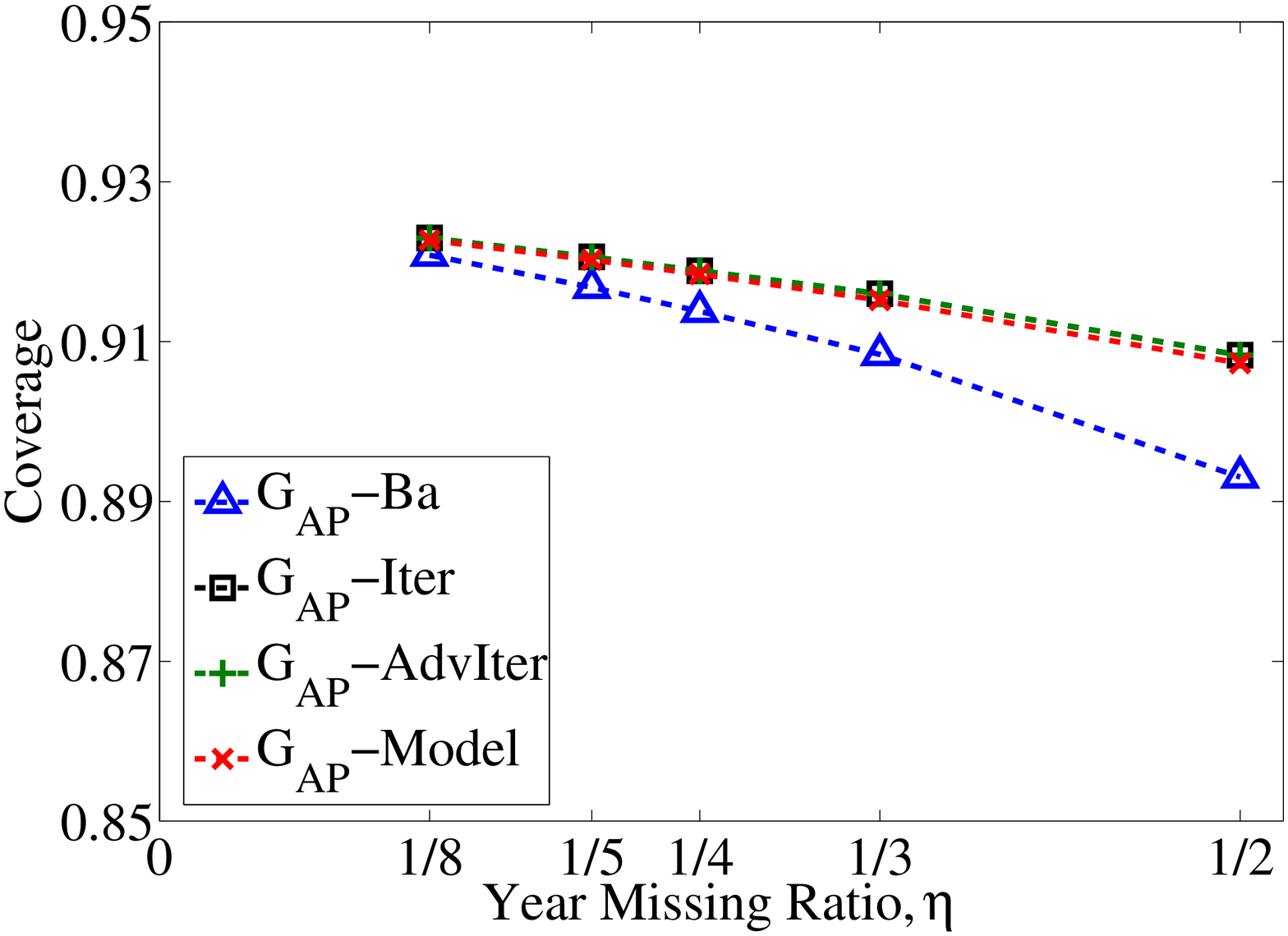}}
    \subfigure[Coverage-DBLP]{
    \label{Fig:GAPCovDBLP}
    \includegraphics[width=1.4in]{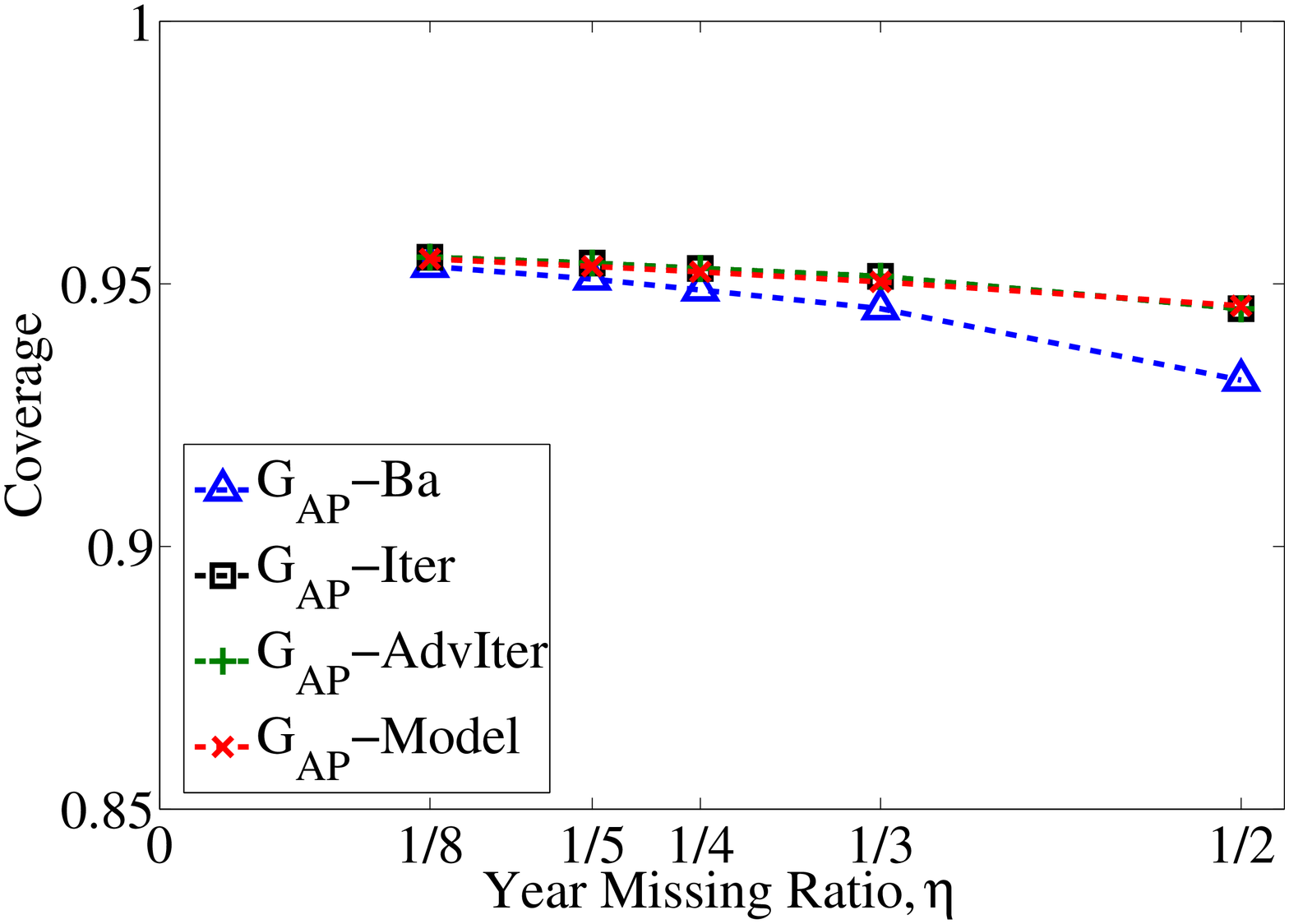}}
    \subfigure[Coverage-APS]{
    \label{Fig:GAPCovAPS}
    \includegraphics[width=1.4in]{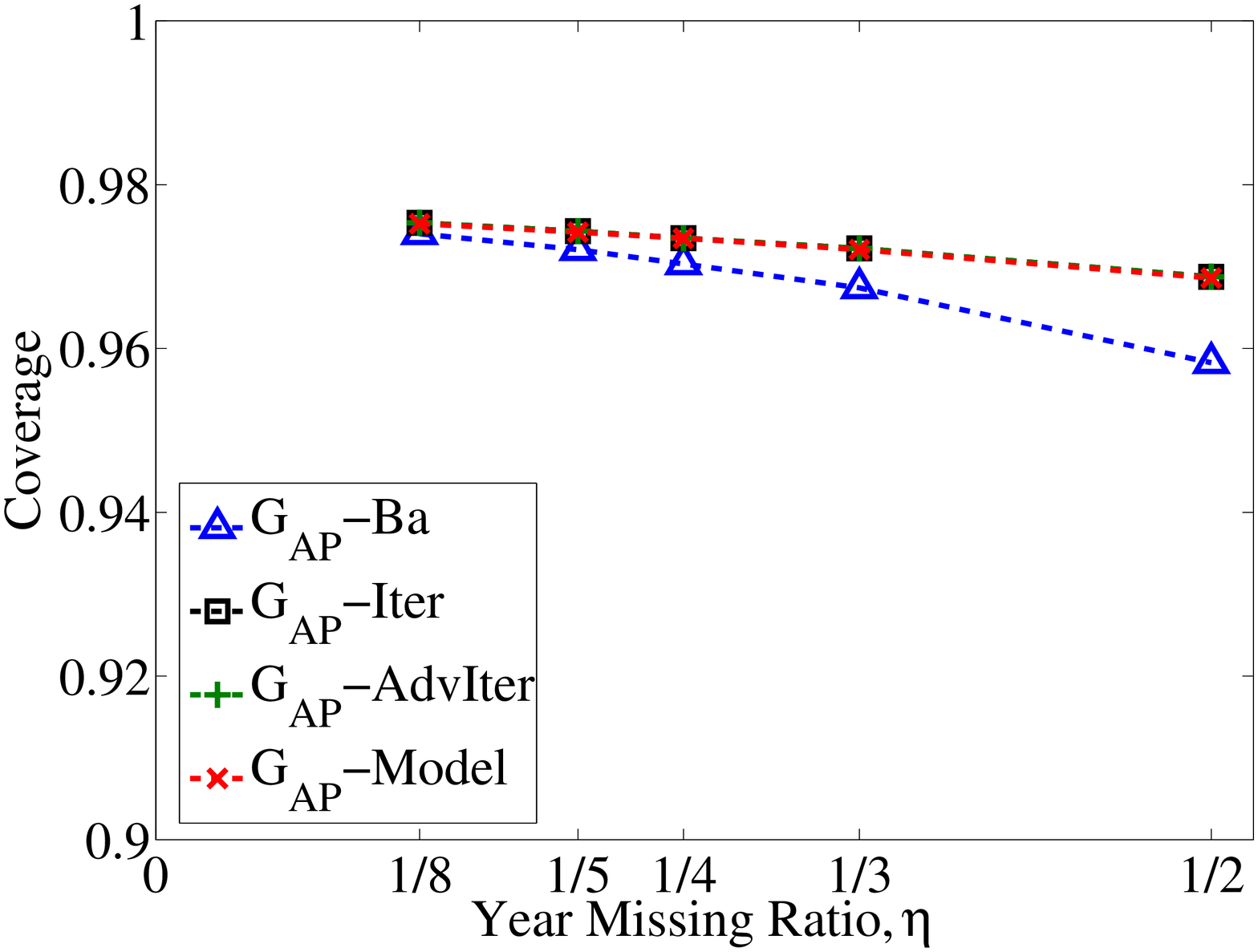}}
    \subfigure[MAE-Libra]{
    \label{Fig:GAPMaeLibra}
    \includegraphics[width=1.4in]{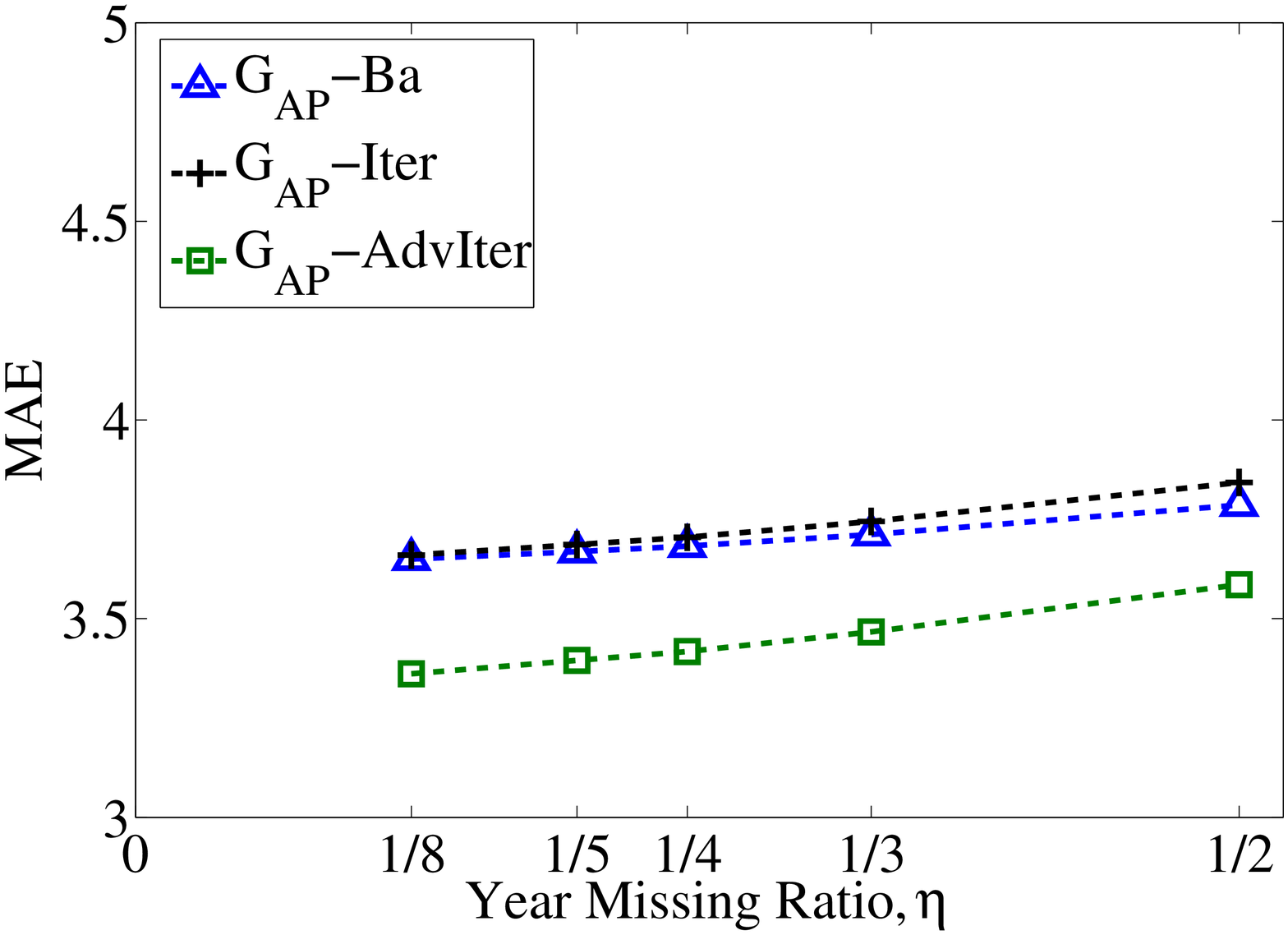}}
    \subfigure[MAE-DBLP]{
    \label{Fig:GAPMaeDBLP}
    \includegraphics[width=1.4in]{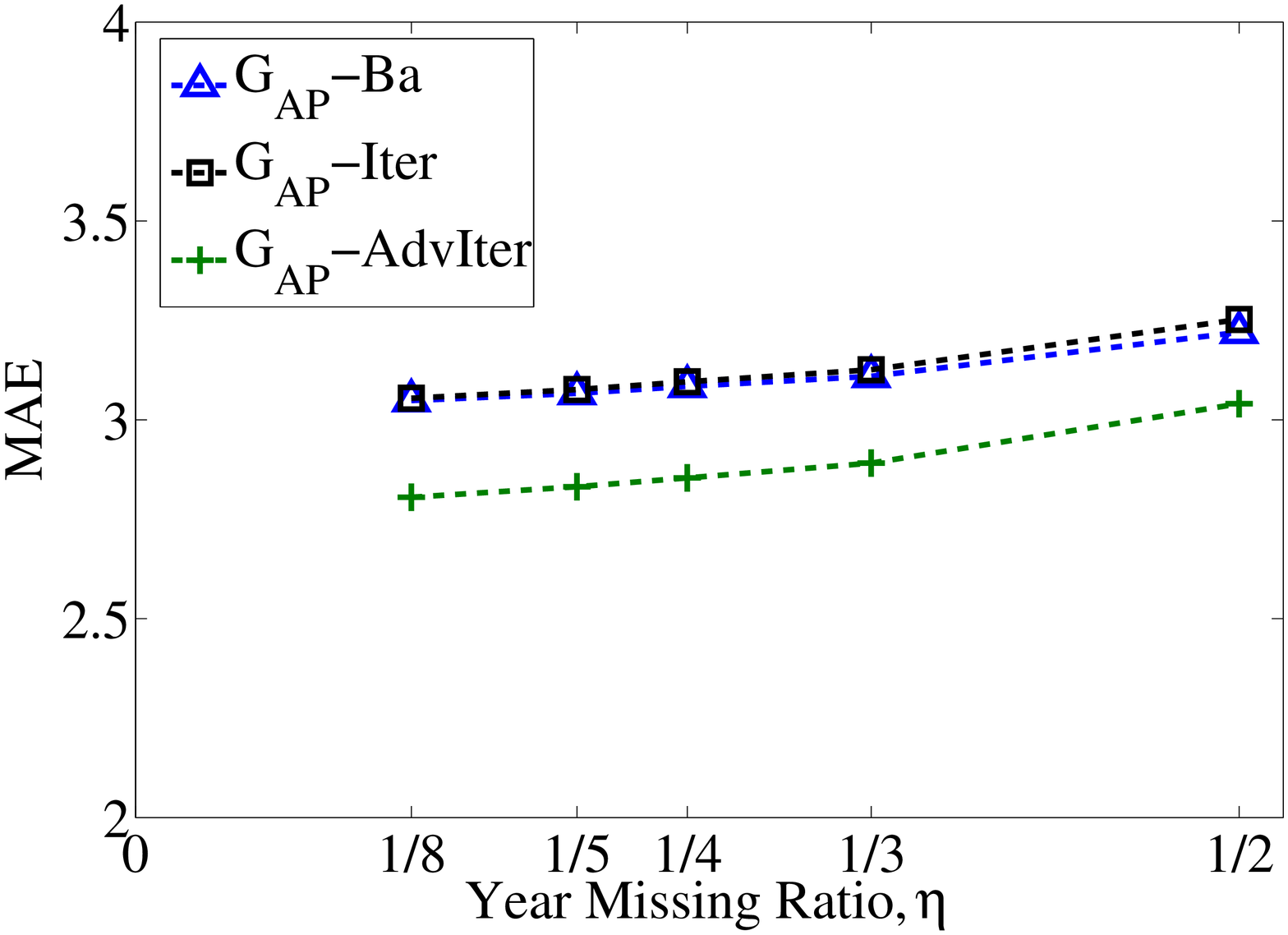}}
    \subfigure[MAE-APS]{
    \label{Fig:GAPMaeAPS}
    \includegraphics[width=1.4in]{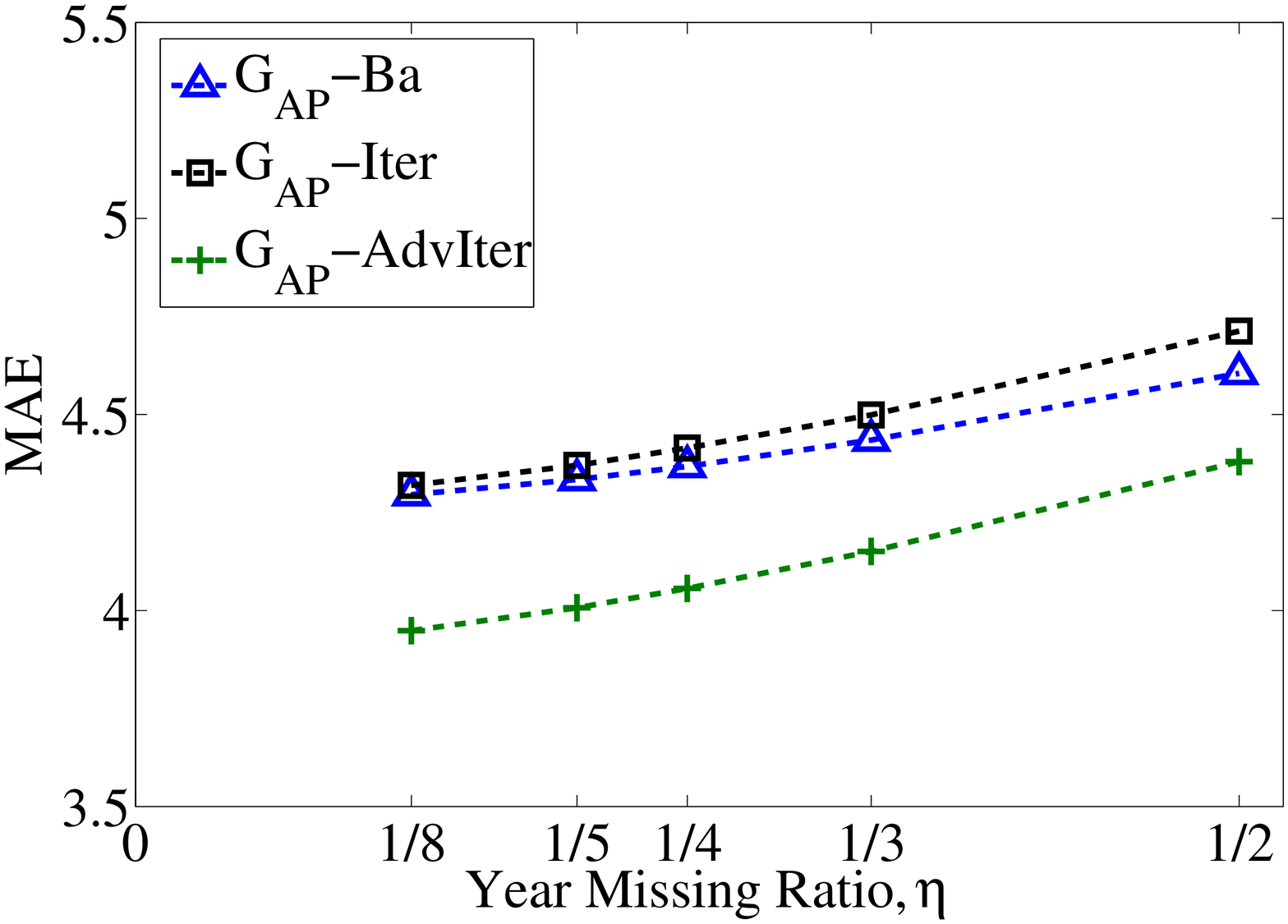}}
    \subfigure[RMSE-Libra]{
        \label{Fig:GAPRmseLibra}
    \includegraphics[width=1.4in]{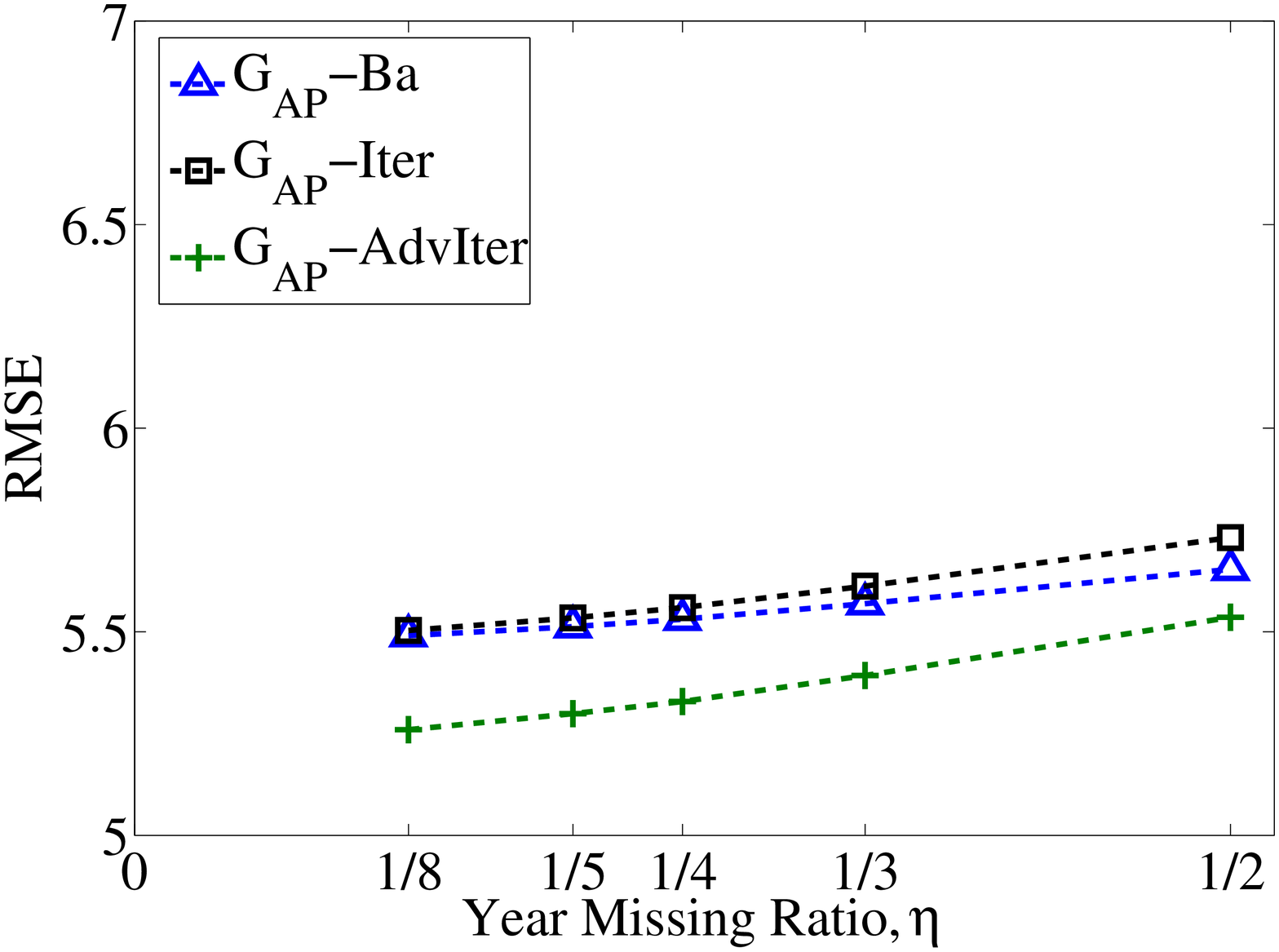}}
    \subfigure[RMSE-DBLP]{
        \label{Fig:GAPRmseDBLP}
    \includegraphics[width=1.4in]{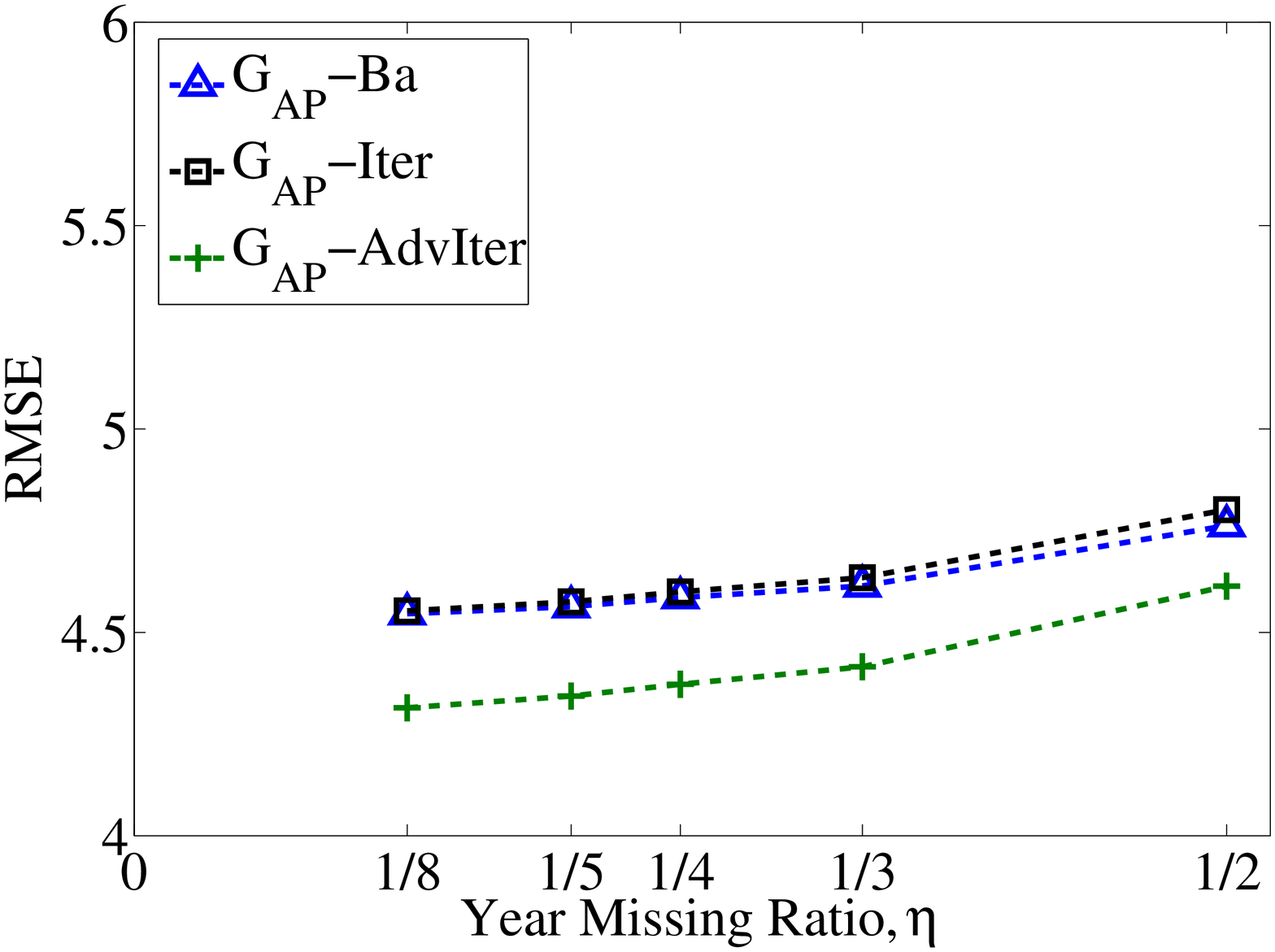}}
    \subfigure[RMSE-ASP]{
    \label{Fig:GAPRmseAPS}
    \includegraphics[width=1.4in]{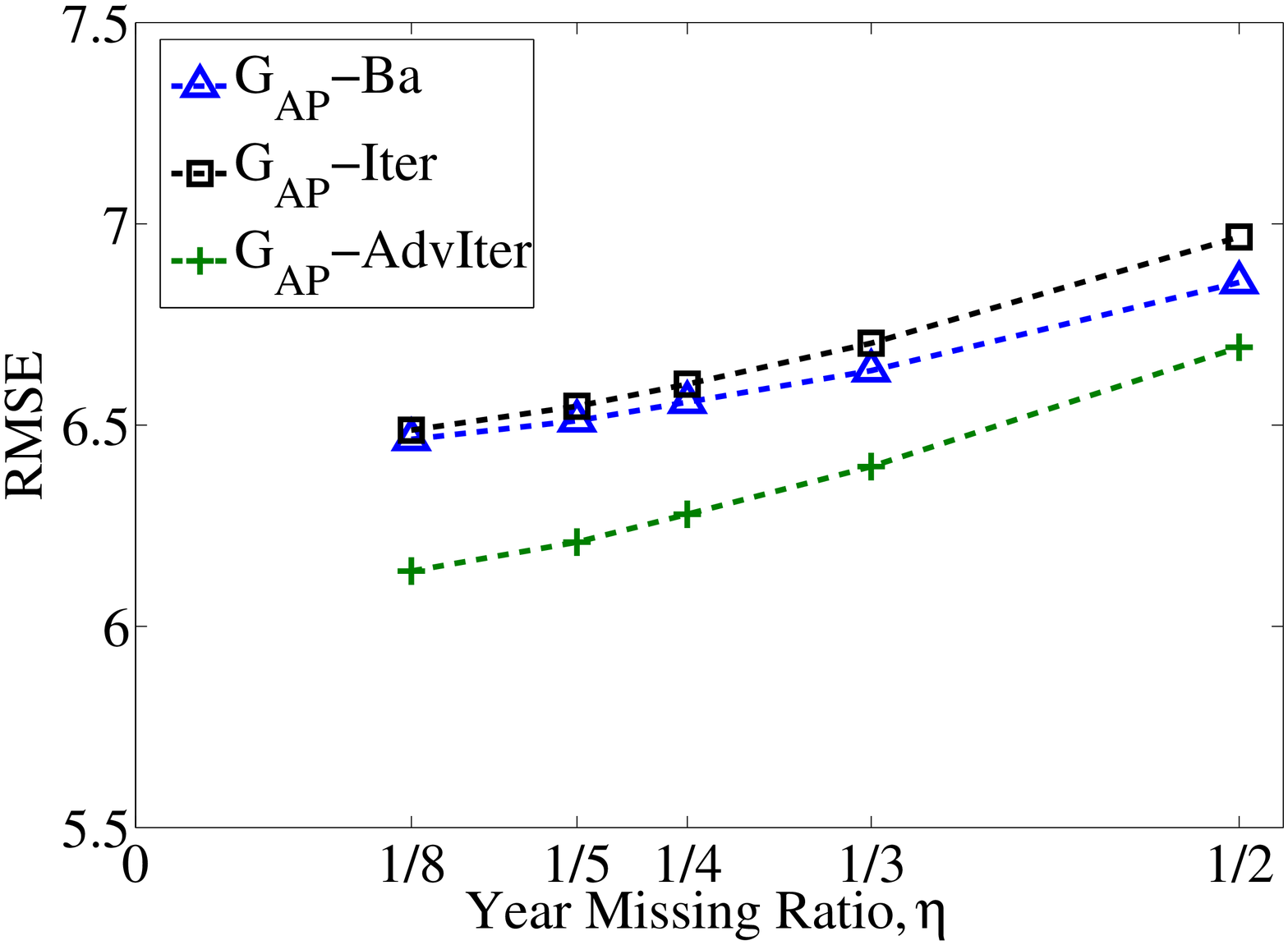}}
    \caption{The coverage, MAE and RMSE of algorithms $G_{AP}$-Ba,
    $G_{AP}$-Iter and $G_{AP}$-AdvIter in paper author bipartite network $G_{AP}$ of the three data sets}
    \label{Fig:GAPLibraDBLPAPS}
\end{figure*}

\begin{enumerate}
\item[1)] In Fig.~\ref{Fig:GAPCovLibra}-\ref{Fig:GAPCovAPS},
the curve of $G_{AP}$-Iter overlaps with that of $G_{AP}$-AdvIter
and has better coverage than $G_{AP}$-Ba. As is discussed
before (Section~\ref{Sec:method}), $G_{AP}$-Iter and
$G_{AP}$-AdvIter utilize the estimation results in the previous
rounds for the later iterations (information propagation) which
leads to the higher coverage results. In addition, the curves of
$G_{AP}$-Iter and $G_{AP}$-Iter match quite well with the expected
value generated by the analytical model.

\item[2)] In Fig.~\ref{Fig:GAPMaeLibra}-\ref{Fig:GAPRmseAPS}, which
concerns estimation accuracy, we find that $G_{AP}$-Iter obtains
worse MAE than $G_{AP}$-Ba. This meets our expectation (in
Section~\ref{Sec:method} that the simple iteration scheme of
$G_{AP}$-Iter spreads inaccuracy during the information
propagation).

\item[3)] It shows that $G_{AP}$-AdvIter performs much better than
the other two in both coverage and accuracy. For all different
$\eta$, $G_{AP}$-AdvIter consistently makes around $10\%$
improvement in MAE measures and $6\%$ in RMSE measures.

\item[4)] If we compare the MAE curves of the three data sets in
Fig.~\ref{Fig:GAPMaeLibra}-\ref{Fig:GAPMaeAPS}, the same algorithm
generates the best MAE on DBLP-Cit data set, the worst on APS data
set and intermediate on Libra data set. This result indirectly
reflects the data quality (on paper-author relationship) of the
three data sets, summarized in Table~\ref{Tab:DataQuaGAP}. As is
widely known that, the original DBLP data set (with no citation
information) is well managed and hence maintains the most complete
and accurate paper-author/ paper-venue relationships~\cite{dblp}.
Libra is an object-level data set, the process of the text-to-object
transfer has been done before we obtain them. Different to the paper
citation links, the APS data set only provides pure text information of
paper-author relationships, therefore, the text-to-object task is
done by ourselves with some simple text-based matching scheme, which
inevitably induces the number of errors in $G_{AP}$. In fact, this
involves several difficult and hot research problems in the
community, for example the Author-Paper Identification Challenge and
the Author Disambiguation Challenge in~\cite{kddcup2013}.

\begin{table}[htb]
\centering
\begin{tabular}{|c|c|}
\hline
\multicolumn{2}{|c|}{MYE performance in $G_{AP}$}\\
\hline    Coverage    & DBLP $\approx$ Libra $\approx$ APS\\
\hline    MAE/RMSE    & DBLP $<$ Libra $<$ APS\\
\hline
\multicolumn{2}{|c|}{Inferred data quality of paper-author relationship}\\
\hline    Completeness & DBLP $\approx$ Libra $\approx$ APS\\
\hline    Correctness  & DBLP $>$ Libra $>$ APS\\
\hline
\end{tabular}
\caption{Summary on data quality of paper-author relationship of
three used datasets inferred from MYE performance in $G_{AP}$.}
\label{Tab:DataQuaGAP}
\end{table}
\end{enumerate}

\subsection{Experiment results for the heterogenous network $G$}
\begin{figure*}[hbt]
    \centering
    \subfigure[Coverage-Libra]{
    \label{Fig:GCovLibra}
    \includegraphics[width=1.4in]{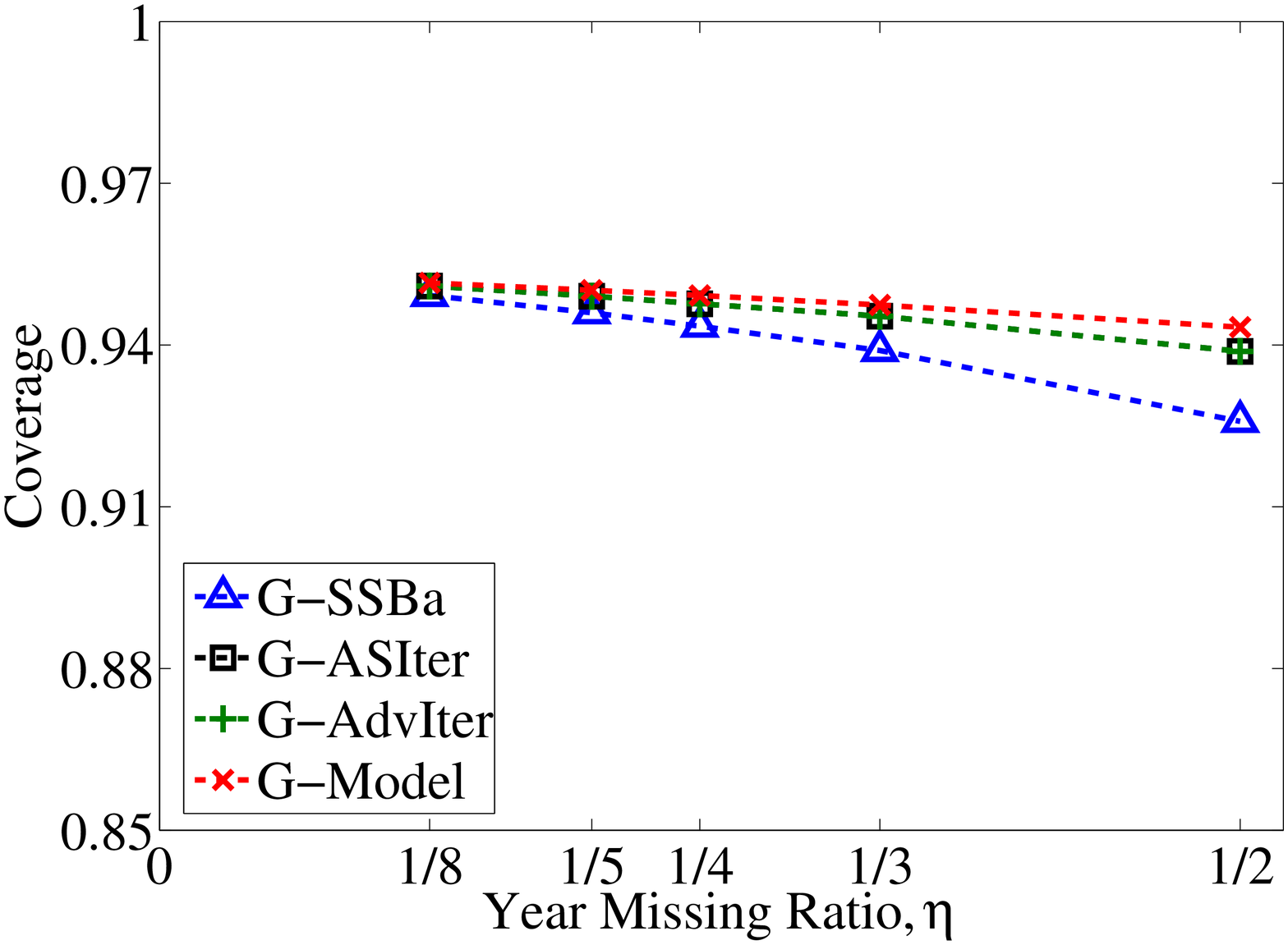}}
    \subfigure[Coverage-DBLP]{
    \label{Fig:GCovDBLP}
    \includegraphics[width=1.4in]{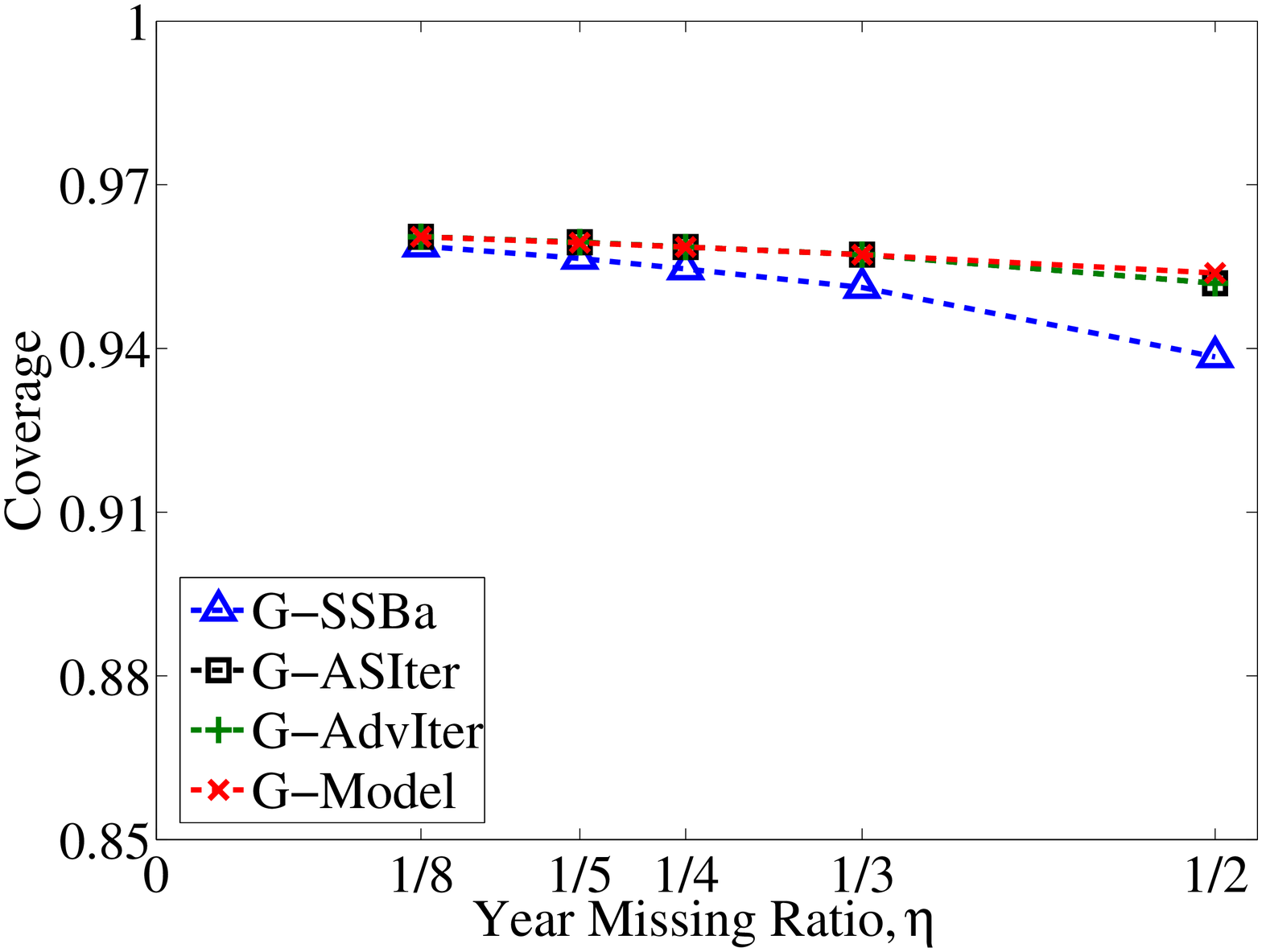}}
    \subfigure[Coverage-APS]{
    \label{Fig:GCovAPS}
    \includegraphics[width=1.4in]{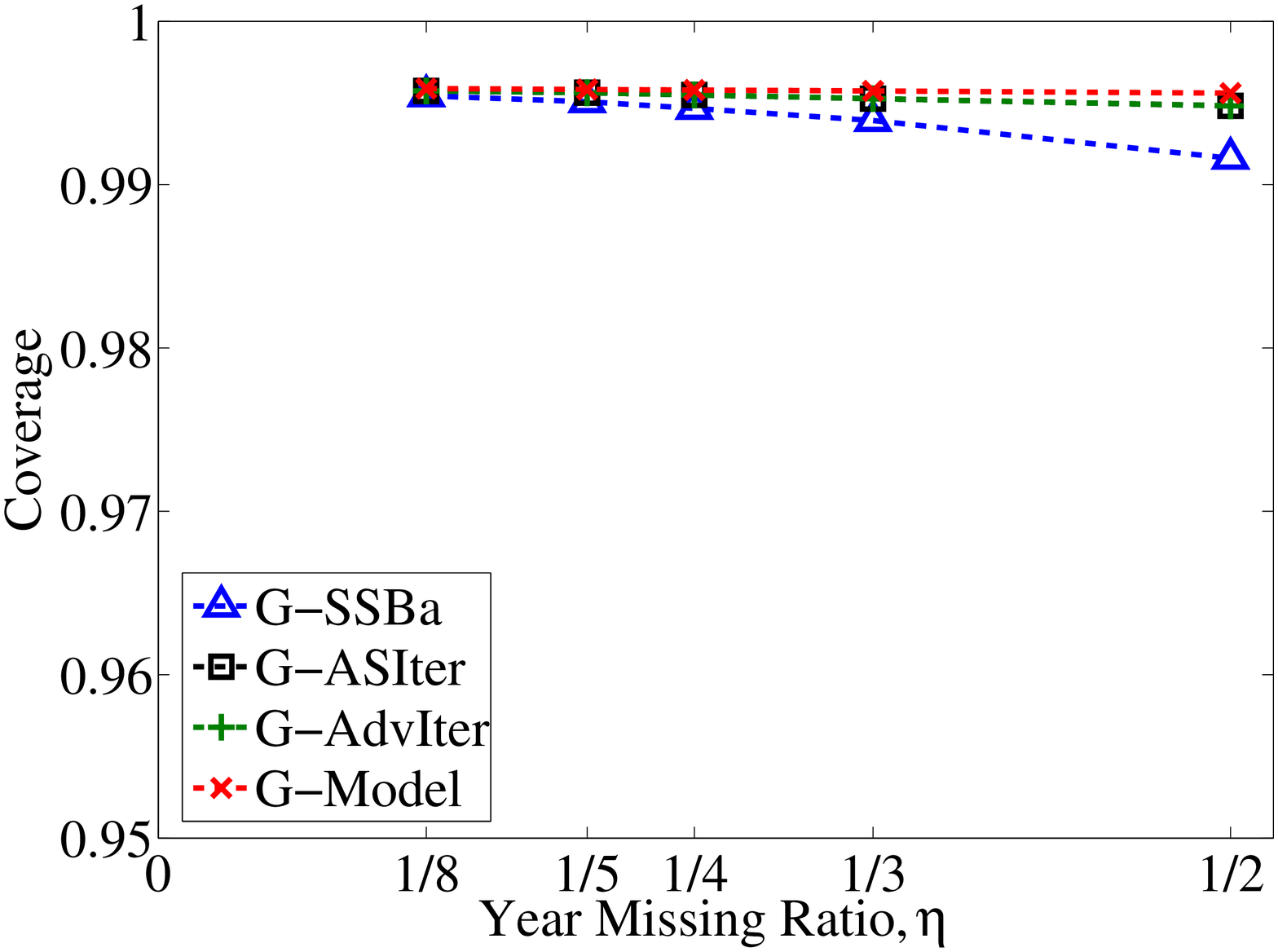}}
    \subfigure[MAE-Libra]{
    \label{Fig:GMaeLibra}
    \includegraphics[width=1.4in]{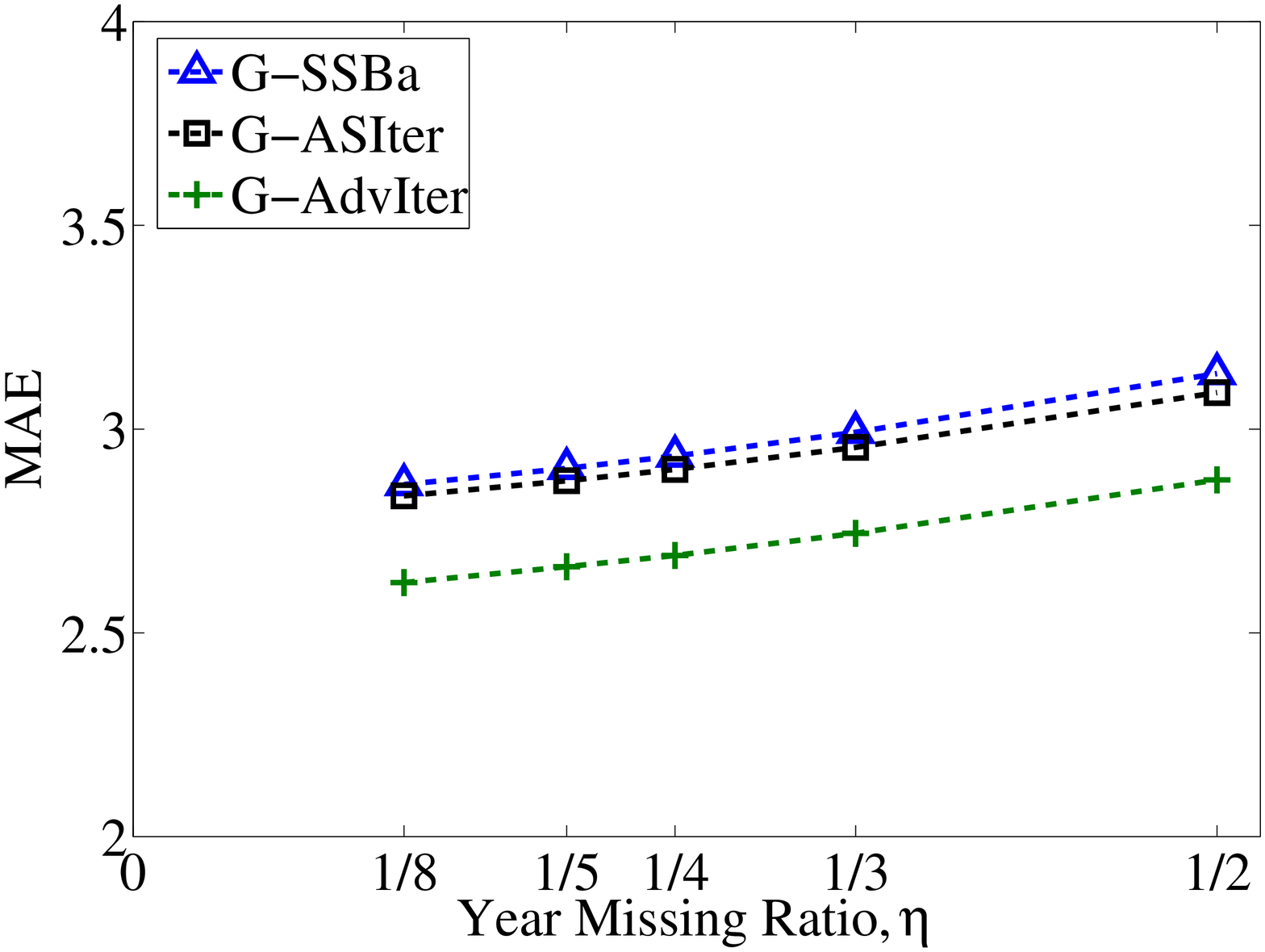}}
    \subfigure[MAE-DBLP]{
    \label{Fig:GMaeDBLP}
    \includegraphics[width=1.4in]{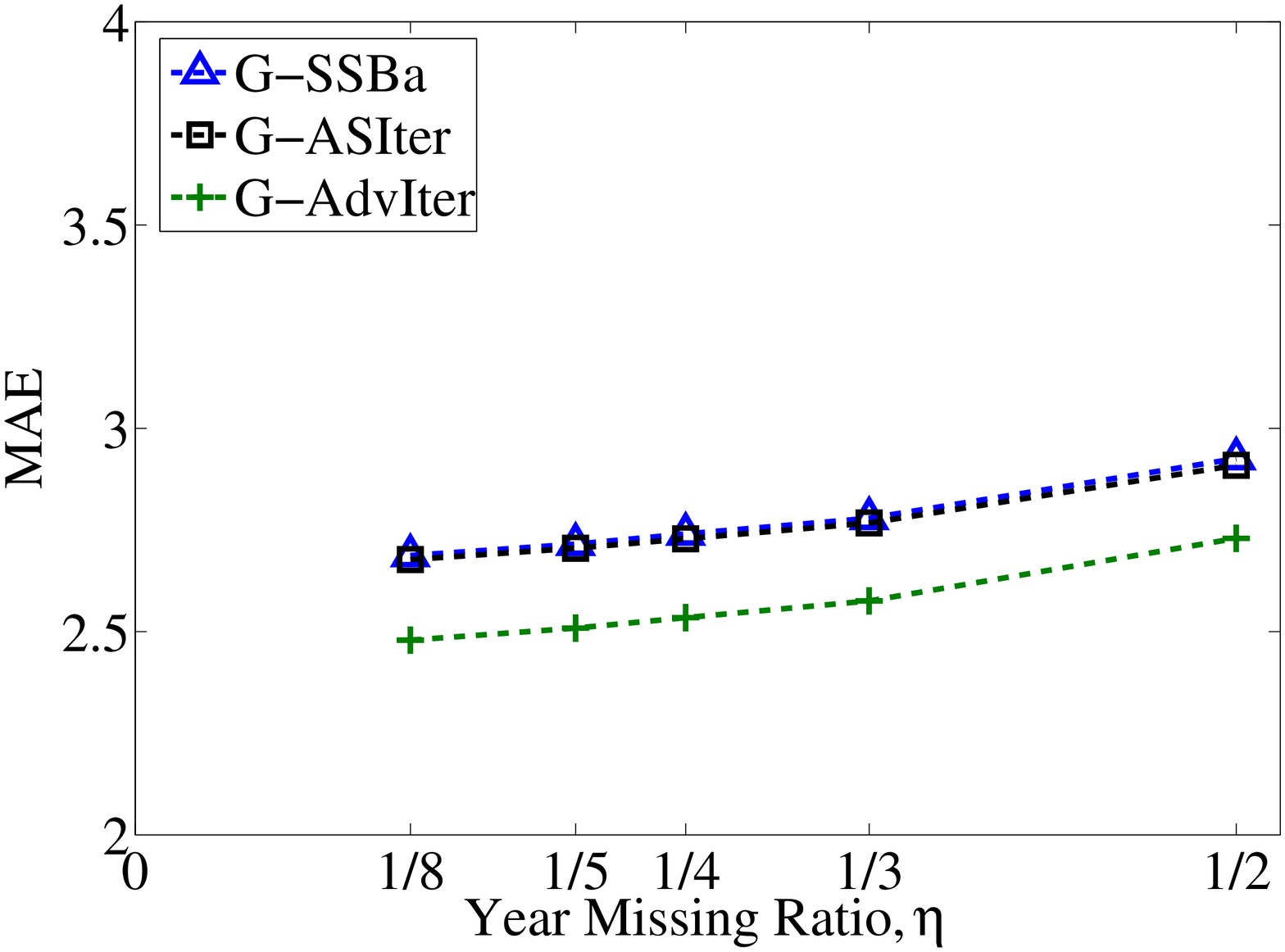}}
    \subfigure[MAE-APS]{
    \label{Fig:GMaeAPS}
    \includegraphics[width=1.4in]{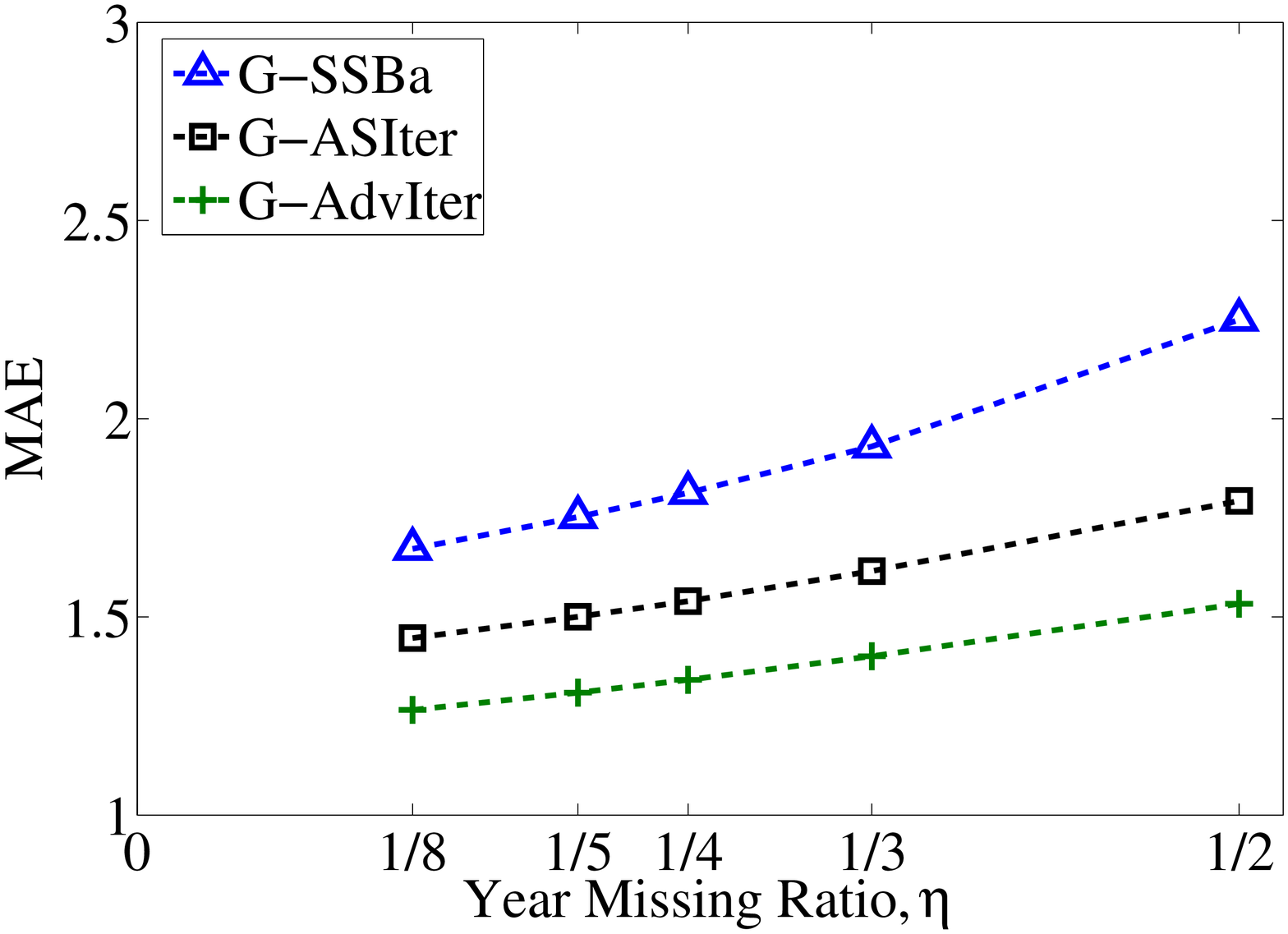}}
    \subfigure[RMSE-Libra]{
    \label{Fig:GRmseLibra}
    \includegraphics[width=1.4in]{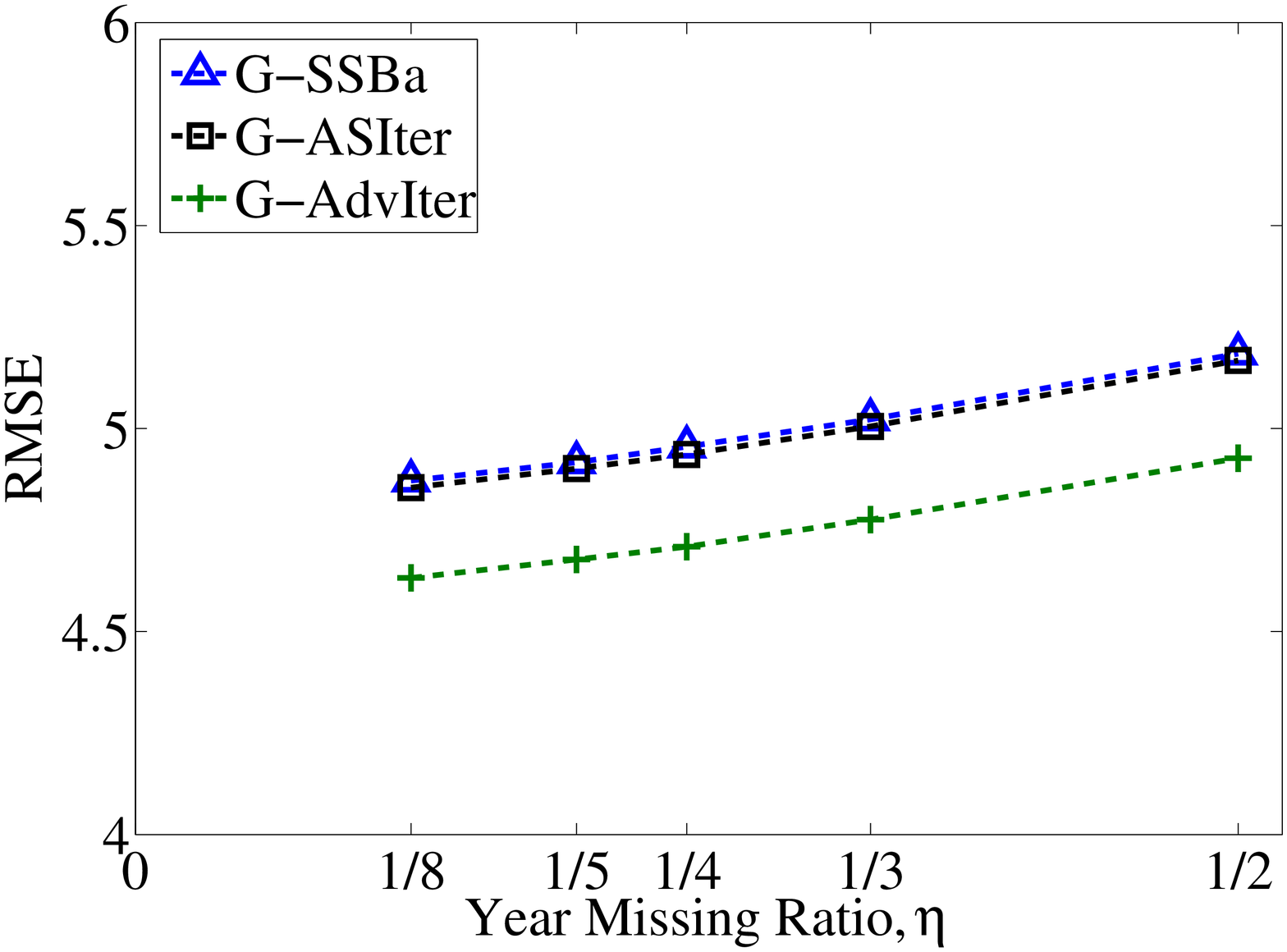}}
    \subfigure[RMSE-DBLP]{
    \label{Fig:GRmseDBLP}
    \includegraphics[width=1.4in]{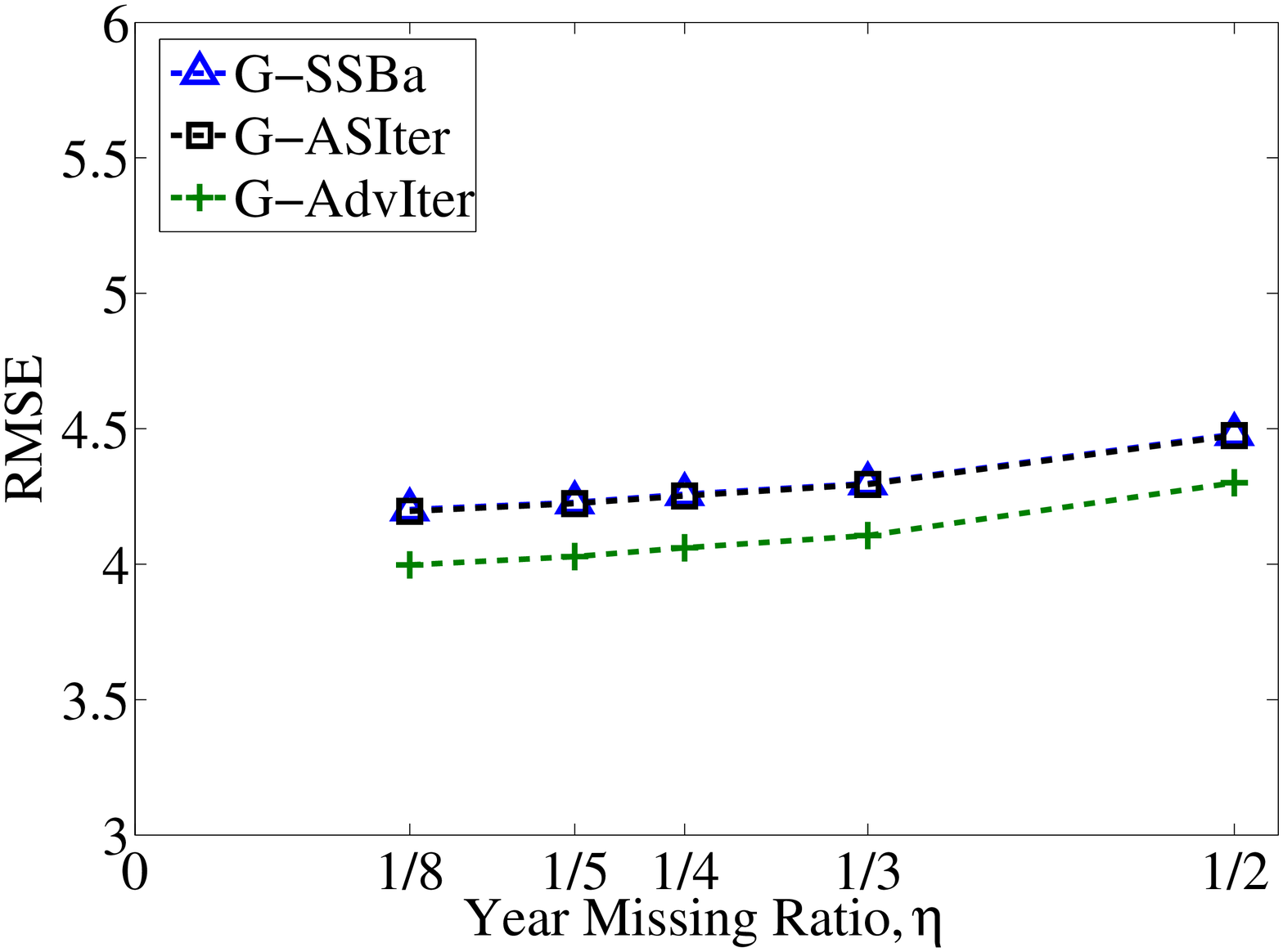}}
    \subfigure[RMSE-APS]{
    \label{Fig:GRmseAPS}
    \includegraphics[width=1.4in]{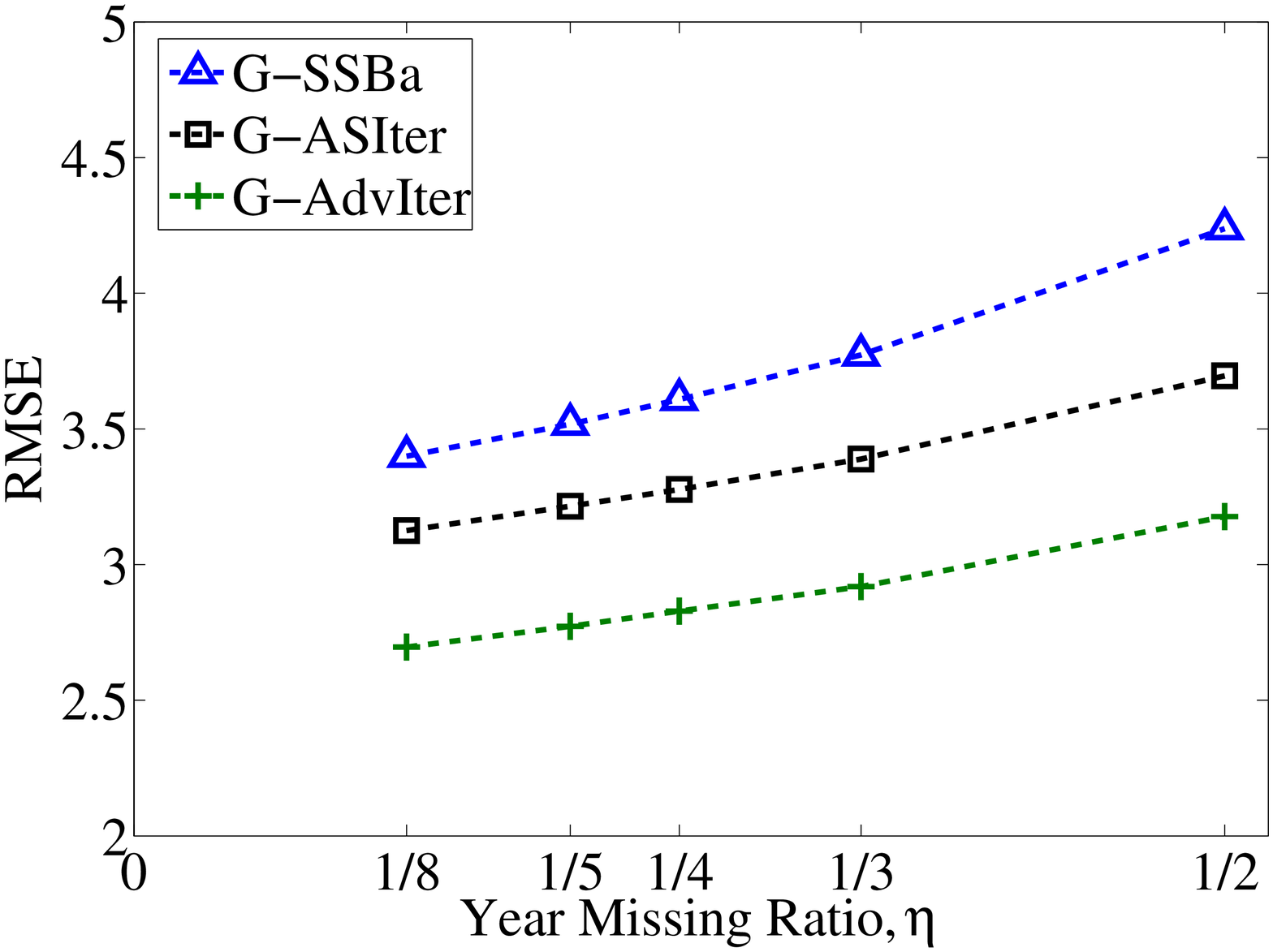}}
    %
    \caption{The Coverage, MAE and RMSE of algorithms $G$-SSBa,
    $G$-ASIter and $G$-AdvIter in the heterogenous network $G$ of the three data sets.}
    \label{Fig:GLibraDBLPAPS}
\end{figure*}

The last set of experiments are conducted on the heterogeneous
network $G = (G_P \cup G_{AP})$ which consists of both the paper
citation network and the paper author bipartite network. The
coverage, MAE and RMSE results of algorithms $G$-SSBa (combination
of $G_P$-SS and $G_{AP}$-Ba), $G$-ASIter (combination of $G_P$-AS
and $G_{AP}$-Iter) and $G$-AdvIter (combination of $G_P$-AA and
$G_{AP}$-AdvIter) are plotted in Figure~\ref{Fig:GLibraDBLPAPS}.

We make three observations according to the results shown in
Fig.~\ref{Fig:GLibraDBLPAPS}:
\begin{enumerate}
\item[1)] All the curves have similar shapes as those in
Fig.\ref{Fig:GPLibraDBLPAPS} and Fig.\ref{Fig:GAPLibraDBLPAPS}, but
the results in Fig.\ref{Fig:GLibraDBLPAPS} have the highest coverage
and smallest MAE and RMSE. This shows the advantage of the
heterogeneous information (both paper citation and paper author
relationship) and the proper combination of the MYE algorithms in
$G_P$ and $G_{AP}$.

\item[2)] In Fig.~\ref{Fig:GCovLibra}-\ref{Fig:GCovAPS},
there appears certain deviations (although milder than those in
Fig.~\ref{Fig:GPCovLibra}-\ref{Fig:GPCovAPS}) from the coverage
curves of $G$-ASIter and $G$-AdvIter to that generated by the
analytical model. This is again due to the overestimation of the
expected number of covered papers by the citation network
information, since $G$-ASIter and $G$-AdvIter are the combinations
from $G_P$-AS and $G_{P}$-AA respectively.

\item[3)] The $G$-AdvIter outperforms the other two for both
coverage and accuracy (with around $8\%$ improvement in MAE and
$5\%$ in RMSE for different $\eta$).
\end{enumerate}

\section{Related Works}\label{Sec:related}
%
%
In network analysis, early studies focused on the structural
characteristics of missing data, e.g.,~\cite{kossinets2006effects}.
\cite{borgatti2006robustness} studied the impact of the measurement
errors on random Erd\H{o}s-R\'{e}nyi networks. A more recent work
by~\cite{wang2012measurement} reclassifies measurement errors,
separating missing data and false data, then analyzes their efforts on
different topology properties of an online social network and a
publication citation network. However, only few works study techniques to
correct measurement errors.

%
Variants of the well-known PageRank~\cite{brin1998anatomy} and
HITS~\cite{kleinberg1999authoritative} algorithms are often used in
social network analysis. \cite{nachenberg2010polonium} use an
iterative Belief Propagation Algorithm to identify malware from a
large scale of files and machines. \cite{zhu2005semi} study the
propagation of two or more competing labels on a graph, using
semi-supervised learning methods.
%

Temporal information is frequently used in topics of an academic
network, e.g.~\cite{chiu2010publish,fu2013asn}. In the research of Academic
Rankings,~\cite{stringer2008effectiveness} find nearly all journals
will reach a steady-state of citation distribution within a
journal-specific time scale, thus they proposed a model for the rank of
paper impacts using citation counts. To solve the tricky problem of
name disambiguation in a digital library,~\cite{tang2012unified}
utilized the multi-hop co-author relationship and its special
property of time-dependence. \cite{wang2010mining} proposed a
time-constrained probabilistic factor graph model to mining the
highly time-dependent advisor-advisee relationship on the
collaboration network.
%

%
The topic of evolution of communities also attracts much attention.
\cite{blei2006dynamic} have used state space models on the natural
parameters of the multinomial distributions to represent the dynamic
evolution of topics. \cite{iwata2010online} developed the continuous
time dynamic model to mine the latent topics through a sequential
collection of documents. ~\cite{gupta2011evolutionary} proposed an
algorithm that integrates clustering and evolution diagnosis of
heterogeneous bibliographic information networks.
\cite{lin2011joint}track the evolution of an arbitrary topic and
reveal the latent diffusion paths of that topic in a social
community. \cite{li2012adding} addressed the community detection
problem by integrating dynamics and communities into the topic
modeling algorithms, and experimented on the Scholarly publications
data set ArnetMiner~\cite{Tang:08KDD}, and recently to generalize the
preferential attachment model with considering the aging factor~\cite{wu2013PAaging}.

Recently, data cleaning on academic social networks receive much
attention. In KDD Cup 2013, the two challenges are the Author-Paper
Identification Challenge or the Author Disambiguation Challenge. For
both challenges, the publishing year information of each paper is
important background knowledge and affect the design of the
algorithms. However, the given data set~\cite{kddcup2013} has a high
\emph{Missing Year Ratio}, $ \eta = \frac{155784}{2257249}\approx
6.90\%$. This is one of the practical examples and usages that
imply the importance of the MYE problems and provide good motivation for this work.

\section{Conclusions}\label{Sec:conclusion}
In this paper, we are dealing with the papers' missing publication
year recovery problem in the Academic Social Network (ASN). We have
considered using three possible networks for estimating missing
years: the paper citation network, the paper author bipartite
network and the heterogenous network (the combination of the
previous two). In each network, we first propose a simple algorithm
which is considered as a benchmark. Next another algorithm involving
information propagation mechanism is proposed. The propagation
mechanism helps to increase the estimation coverage ratio. Finally,
an advanced propagation based algorithm is proposed, and in each of
the three networks the advanced algorithm outperforms other
algorithms and achieves at least an $8\%$ improvement on MAE and $5\%$
on RMSE. In addition, the coverage achieved by the advanced
algorithms well matches the results derived by the analytical model.

\bibliographystyle{spbasic}      
\bibliography{paper_mye_ref}   

%
%

\end{document}